\documentclass[10pt,aps,prl,reprint,twocolumn,superscriptaddress,floatfix,showpacs,longbibliography]{revtex4-2}
\usepackage[utf8]{inputenc}
\usepackage[T1]{fontenc}
\usepackage{graphicx}
\usepackage{tabularx}
\usepackage[usenames,dvipsnames,table]{xcolor}
\usepackage[version=3]{mhchem}
\usepackage{braket}
\usepackage{upgreek}
\usepackage{hyperref}
\usepackage{amsmath, amssymb}
\usepackage[normalem]{ulem}
\usepackage{soul}
\usepackage{bbold}
\usepackage{multirow}

\definecolor{mark}{rgb}{0.85, 0.9, 1}

\hypersetup{hidelinks}
\sethlcolor{mark}

\begin{document}

\title{Intrinsic dimension of quantum sampling data from nearest-neighbor Hamming distances}

\author{P. S. Golubev}
\affiliation{Theoretical Physics and Applied Mathematics Department, Ural Federal University, Ekaterinburg 620062, Russia}
\author{I. A. Iakovlev}
\affiliation{Theoretical Physics and Applied Mathematics Department, Ural Federal University, Ekaterinburg 620062, Russia}
\affiliation{Russian Quantum Center, Skolkovo, Moscow 121205, Russia}
\author{V. V. Mazurenko}
\affiliation{Theoretical Physics and Applied Mathematics Department, Ural Federal University, Ekaterinburg 620062, Russia}
\affiliation{Russian Quantum Center, Skolkovo, Moscow 121205, Russia}

\begin{abstract}
Quantum data can be obtained from a diverse range of sources, including direct measurements from noisy quantum processors, cold-atom simulators, and classical approximations such as variational neural-network states. However, our ability to characterize these systems is fundamentally limited, as the available measurement data are often sparse compared to the exponentially large Hilbert space of the system. To address this, we propose using the expected nearest-neighbor Hamming distance calculated for a set of unique bitstrings as a geometric descriptor revealing a distinct power-law behaviour. Crucially, under a locally uniform probability density approximation, we relate the empirical power-law exponent to an effective intrinsic dimension of the sampled measurement support in a given measurement basis and clarify the finite-sample regimes where this correspondence holds. Through representative benchmarks on experimental and simulated datasets, we demonstrate that these power-law parameters capture key geometric features of measurement outcomes, provide signatures of quantum phase transitions and critical dynamics from limited sampling data, and diagnose the optimization accuracy of variational neural-network states. This enables the analysis of completely different quantum experiments within a single framework. Furthermore, we demonstrate the versatility of our approach on classical high-dimensional data, showing that the power-law parameters can successfully discriminate different classes of handwritten digits from the MNIST dataset.  
\end{abstract}
\maketitle

\section{Introduction} 
Quantum state measurement is an integral part of any quantum experiment, including quantum computing, where the final wave function encodes a solution of a problem. The resulting basis states, represented as bitstrings (for instance, '01000..1', '11010..1'), provide only limited information regarding the underlying quantum state, as they are typically few in number compared to the dimensionality of the Hilbert space. This makes extracting the maximum amount of useful information from quantum measurements a nontrivial problem that lacks a universal solution. Numerous measurement-based approaches for characterizing quantum states developed to date mainly focus on estimating standard inter-qubit correlation functions and entropies of different types \cite{Heyl_random, classical_shadow1, classical_shadow2, MINE, QMINE}. Alternative approaches  \cite{dissimilarity, struct-complexity, dissimilarity_sign_structure, Scalettar, Khatami} treat bit-string arrays as datasets, revealing patterns and quantifying correlations between different samples, effectively abstracting away the quantum origin of the input. Here, various metrics from data science can be applied, and subsequently leveraged as hash functions for the underlying quantum states, facilitating their efficient discrimination. 

A primary procedure in processing bit-string arrays for some high-level algorithms mentioned above is the calculation of the Hamming distance \cite{Hamming}, which quantifies the number of bits by which two samples differ. For instance, within the wave-function network approaches \cite{Dalmonte1, boson_sampler_our, Boson_sampler_Rubtsov, Heyl, Dalmonte3} the minimal Hamming distance averaged over a bit-string array determines the structure of the graph subsequently used for analysis of the quantum state. In certain situations, the Hamming distance on its own facilitates the solution of a problem where error correction \cite{Frank} and discrimination of non-trivial non-equilibrium quantum states \cite{DTC, DTCour} serve as prominent examples. While the Hamming distance metric is extensively applied in quantum computing, its behavior is still poorly understood. 

In this work, we characterize state spaces of notable quantum states at the level of projective measurements and reveal a previously unreported  feature of the Hamming distance. Namely, for quantum sampling data recorded in a given measurement basis, the average nearest-neighbor Hamming distance follows a power-law scaling with the unique-bitstring subset size. This scaling behavior reveals the connectivity within the state space and, as we demonstrate from nearest-neighbor statistical arguments, is related to the fundamental concept of intrinsic dimension. The corresponding power-law parameters serve as the effective descriptors of the quantum state and enable an efficient state discrimination of large-scale wave functions with a limited number of measurements. In turn, the analysis of the antiferromagnetic $J_1 -J_2$ model solutions obtained with the attention-based vision transformer wave function facilitates detecting phase boundaries and estimating accuracy of the neural network approach at different points of the phase space. As an example of non-equilibrium problem we consider the quench dynamics of spin-glass systems simulated on D-Wave quantum processors. Furthermore, a real-world problem that is discrimination of the MNIST digits by their intrinsic dimensions can be likewise solved with our approach.

\begin{figure}[hbt!]
    \includegraphics[width=\linewidth]{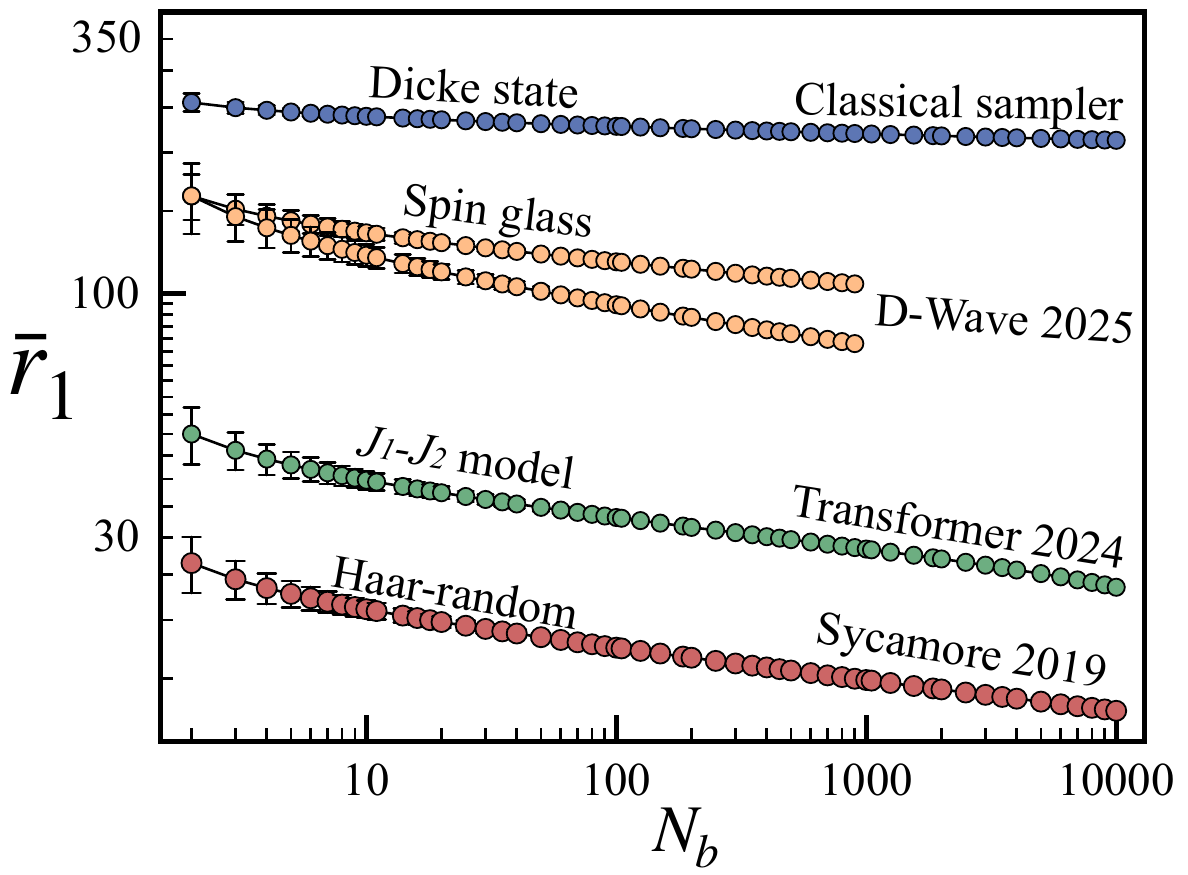}
    \caption{Expected minimal Hamming distance functions calculated for a few notable quantum states (from top to bottom): the classically-sampled 512-qubit Dicke state with 256 excitations; 324-qubit spin-glass states simulated on the D-Wave quantum annealer \cite{DWave2025} with the square lattice topology and quench times of 7 (the upper yellow curve) and 20 (the lower yellow curve) ns; ground state of the frustrated antiferromagnetic $J_1 - J_2$ model optimized with transformer-based neural network quantum state approach \cite{DeepViT} on $10 \times 10$ square lattice at $J_2 = 0.5$; 53-qubit Haar-random state imitated on Sycamore quantum processor \cite{Sycamore}.}
    \label{fig1}
\end{figure}

\section{Method} 
To characterize an $N$-qubit quantum system described by a wavefunction $\ket{\Psi}$, one performs a series of projective measurements in a specific basis. From these measurements, we collect a total of $N_b^{\text{max}}$ unique bitstrings, where the $i$-th configuration is represented as ${\bf x}_i = \{x_i^k\}_{k=1, \dots, N}$ with $x_i^k \in \{0,1\}$. Depending on the particular source of measurement data, this total count can vary from $10^3$ for the D-Wave quantum annealer to $10^6$ for the Google Sycamore device. Since each sample in this preprocessed record is unique, the probabilities of individual basis states and related frequency-dependent 
quantities cannot be extracted directly from the data. 

As the next step, we randomly subsample $N_b$ configurations from the entire set of $N_b^{\text{max}}$ unique outcomes and calculate the mutual Hamming distances between them, $d({\bf x}_i, {\bf x}_j) = \sum_{k=1}^N |x^k_i - x^k_j|$, where $i,j = 1, \dots, N_b$ and $i \neq j$. For each bitstring within this subsample, we determine the distance to its first nearest neighbor.

To ensure statistical robustness and eliminate sample-dependent fluctuations, we repeat this generation process to construct several independent datasets of size $N_b$. We compute the mean minimal Hamming distance for each dataset and subsequently perform an ensemble average over the resulting values. In this way, we define the expected nearest-neighbor distance function as follows:
\begin{equation}
\bar{r}_1(N_b) = \left\langle \frac{1}{N_b} \sum_{i=1}^{N_b} \min_{j \neq i} d({\bf x}_i, {\bf x}_j) \right\rangle,
\label{Hamdist}
\end{equation}
where $\langle \dots \rangle$ denotes the subsampling average evaluated over $N_{\mathrm{sub}}$ randomly drawn subsets of the size-$N_b$ datasets. It is important to distinguish this internal subsampling average in Eq.~\eqref{Hamdist} -- performed over $N_{\mathrm{sub}}$ random subsets drawn from a single fixed measurement record to suppress combinatorial selection fluctuations -- from outer ensemble averages over independent experimental runs or problem realizations (instances) reported in later sections. Because each subsample already averages nearest-neighbor distances over $N_b$ configurations, the resulting $\bar{r}_1(N_b)$ curves converge rapidly with $N_{\mathrm{sub}}$, so that $N_{\mathrm{sub}} = 50$ repetitions are typically sufficient across all our benchmarks.

Usually, the average minimum distance $\bar{r}_1$ calculated at a given $N_b$ is to play a supporting role being used as the cutoff radius when constructing wave function networks \cite{Dalmonte1, Heyl, Dalmonte3}. Here we show that the $\bar{r}_1(N_b)$ scaling provides sufficient information for identifying and discriminating quantum states of different complexity, probing signatures of phase boundaries, as well as analyzing high-dimensional classical datasets.

Fig.~\ref{fig1} displays the $\bar{r}_1 (N_b)$ dependencies in a log-log scale for several notable large-scale quantum states, for which measurement data are either publicly available from real quantum devices or can be  generated on classical computers. Despite the different origin of underlying quantum data, the resulting Hamming functions calculated with measurements in the $\sigma^z$ basis reveal similar characteristic patterns. Namely, a rapid non-linear decrease with large standard deviations at small values of $N_b$ is replaced by almost linear behaviour with a slope specific to each wave function. The linear part of the $\bar{r}_1 (N_b)$ function in the log-log scale can be fitted with
\begin{eqnarray}
\label{power-law}
\bar{r}_1 (N_b) = A N_b^{-\alpha}.
\end{eqnarray}
This motivates using the power-law parameters $A$ and $\alpha$ as basis-dependent geometric descriptors to characterize the underlying structure of the sampled bitstrings in a computational basis. While the coefficient $A$ in Eq.~\eqref{power-law} corresponds to the value of $\bar{r}_1 (N_b)$ at $N_b=1$, the minimal number of bitstrings for which the Hamming distance function can be defined is two. Nevertheless, the coefficient $A$ demonstrates a high sensitivity to variations of a given quantum state, which is particularly useful for efficient state discrimination and can be formally obtained via the $N_b \rightarrow 1$ extrapolation of $\bar{r}_1(N_b)$. In turn, the exponent $\alpha$ determines the scaling rate of finding the nearest neighbors in the state space as the sample set increases, which turns out to be slow, with $\alpha \in [0.1, 0.35]$ for the quantum states we consider in this work. In Appendix~A, we present the additional details regarding the role of bit-string repetitions and the selection of fitting windows.

\subsection{Origin of the power law}
To establish a physical interpretation of the empirical power-law parameters $A$ and $\alpha$, we have reviewed the foundational literature dedicated to nearest-neighbor statistics across various scientific domains. Remarkably, the theoretical roots 
of our approach date back to 1909 with the pioneering work by Paul Hertz \cite{Hertz}, who first derived the exact nearest-neighbor distance distribution to characterize the spatial layout of molecules in an ideal gas. While Hertz originally focused on three-dimensional physical systems, in 1954 this theoretical framework was adapted by biologists \cite{Clark1954}  to evaluate the spatial dispersion of plant populations on a two-dimensional plane, and was eventually generalized to arbitrary $d$-dimensional space by Clark and Evans \cite{Clark1979} in 1979. More specifically, for a random uniform dispersion of independent points in a $d$-dimensional space, the mathematical expectation of the distance to the first nearest neighbor in an ideal gas system ($\bar{r}_{1}^{\text{(gas)}}$) is given by:
\begin{equation}
\label{gas}
\bar r_{1}^{\text{(gas)}} = 
\frac{[\Gamma(\frac{d}{2}+1)]^{1/d} \Gamma(\frac{1}{d}+1)}
{\sqrt{\pi}} \rho^{-\frac{1}{d}},
\end{equation}
where $\Gamma(\cdot)$ is the Euler Gamma function. Here, $\rho$ is the uniform spatial density of the gas \cite{Hertz} or plants \cite{Clark1979}. 

Eq.~\eqref{gas} establishes a direct connection between the nearest-neighbor distance and the embedding system's dimension. 
These ideas were later found to be extremely useful for discovering 
low-dimensional manifold representations of high-dimensional data \cite{Pettis1979, Levina2004}. In particular, Pettis et al.~\cite{Pettis1979} considered a dataset $\{{\bf x}_1, {\bf x}_2, \dots, {\bf x}_n\}$ of $n$ independent samples and, assuming that the underlying probability density is locally constant within the small hypersphere containing the first nearest neighbor, derived the following expression for the average nearest-neighbor distance 
\begin{equation}
\label{pettis_formula}
\bar r_1^{\text{(data)}} \approx 
 C_{\rm geom} (\frac{1}{n} \sum_{i=1}^{n} [p({\bf x}_i)]^{-\frac{1}{d}})  n^{-\frac{1}{d}},
\end{equation}
where $C_{\rm geom} = \frac{[\Gamma(\frac{d}{2}+1)]^{1/d} \Gamma(\frac{1}{d}+1)} {\sqrt{\pi}}$, $d$ is the intrinsic dimension of the underlying manifold, and $p({\bf x}_i)$ denotes the true, unknown local sampling density at ${\bf x}_i$, which is an intrinsic property of the data distribution and does not depend on the sample size $n$. Consequently, when $n$ varies, the sample-averaged factor $\frac{1}{n}\sum_{i=1}^n [p({\bf x}_i)]^{-1/d}$ remains approximately constant (determining the prefactor $A$), so that Eq.~\eqref{pettis_formula} directly yields the power-law dependence $\bar{r}_1^{\text{(data)}} \propto n^{-1/d}$ without requiring explicit knowledge of $p({\bf x}_i)$. If one is going to approximate the true, unknown probability density, this can be done using the nearest-neighbor density estimator $\hat{p}({\bf x}_i) = \frac{1/n}{V({\bf x}_i)}$, where $1/n$ is the elementary probability weight per sample and $V({\bf x}_i) = V_d [r_1({\bf x}_i)]^d$ is the local hypersphere volume with $V_d = \pi^{d/2} [\Gamma(d/2+1)]^{-1}$. To reproduce the basic property of $p({\bf x}_i)$, namely its independence of the sample size $n$, at the level of $\hat{p}({\bf x}_i)$, the local neighbor volume must scale as $V({\bf x}_i) \propto 1/n$.

For continuous Euclidean distributions, Eq.~\eqref{pettis_formula} immediately yields $\bar{r}_1(n) = A n^{-\alpha}$ with $\alpha = d^{-1}$ and $A = C_{\rm geom}\,\frac{1}{n}\sum_{i=1}^n [p({\bf x}_i)]^{-1/d}$. Thus, the coefficient $A$ is functionally linked to the averaged inverse unknown probability density of the data. This means that samples with the smallest $p({\bf x}_i)$, which correspond to maximally sparse areas in the state space, provide the largest contribution to $A$. Moreover, because the underlying density $p({\bf x}_i)$ is independent of the sample size $n$, the prefactor $A$ remains invariant as $n$ increases. The validity of this continuous-space formulation is directly verified in Appendix~C on the MNIST dataset using the $L_2$ norm to calculate distances.

\subsection{Details of mapping onto discrete datasets}
We have shown numerically that the power law in Eq.~\eqref{power-law}, derived from Pettis's continuous-space consideration, also holds for bitstring arrays obtained from quantum state measurements (Fig.\ref{fig1}). To complete this theoretical picture, we extend the derivation of the average nearest-neighbor distance to finite, discrete ensembles of measurement outcomes by mapping the continuous formalism onto the discrete binary hypercube $\{0,1\}^N$ equipped with the integer-valued $L_1$ Hamming metric. Formally, the dependence of the nearest-neighbor distance on the effective dimension in the continuous-space formulation, Eq.~\eqref{pettis_formula}, hinges on the geometric scaling between neighborhood volume and radius, $V(r) = V_d r^d$. Establishing an analogous relation is therefore essential in the discrete case.

On the $N$-dimensional hypercube $\{0,1\}^N$, the maximum Hamming distance is bounded by the total number of bits ($r \leq N$), and the volume of a full Hamming ball grows combinatorially as $V_{\rm H}(r) = \sum_{k=0}^r \binom{N}{k}$ (approaching $N^r/r!$ for $r \ll N$). However, when the sampled bitstrings belong to a structured subset $\mathcal{M} \subset \{0,1\}^N$ of effective dimension $m$, bit correlations partition each string into $m$ effective subgroups (or blocks) of generally unequal sizes $\ell_1, \dots, \ell_m$ ($\sum_{j=1}^m \ell_j = N$). Within each block, bits vary coherently along a one-dimensional chain of local configurations parameterized by an integer coordinate $z_j$ (such that the intra-block Hamming distance between two states is $|\Delta z_j|$), reducing the total Hamming distance to the $m$-dimensional $L_1$ distance $\sum_{j=1}^m |\Delta z_j|$. 

To count all configurations in $\mathcal{M}$ within Hamming radius $r$ of a central bitstring, we group them by the number $k$ of active blocks that differ from the center and thus yield a nonzero contribution to the Hamming distance. Since $k$ cannot exceed either the total number of blocks $m$ or the available bit-flip budget $r$ (as each active block contributes at least one bit flip), it is bounded by $0 \leq k \leq \min(m, r)$. For a given $k$, there are $\binom{m}{k}$ ways to select the active blocks, $2^k$ choices for the direction of change within each active block, and $\binom{r}{k}$ ways to distribute up to $r$ bit flips across these $k$ blocks. Summing over $k$ yields the discrete $L_1$-ball volume~\cite{Golomb1970}
\begin{equation}
V_m(r) = \sum_{k=0}^{\min(m, r)} 2^k \binom{m}{k} \binom{r}{k}.
\end{equation}
When the radius exceeds the effective dimension ($r > m$), the upper summation limit simplifies to $\min(m, r) = m$, and isolating the leading $k = m$ term yields the asymptotic power-law scaling
\begin{equation}
V_m(r) = \frac{2^m}{m!}\, r^m \left[ 1 + \mathcal{O}\!\left(\frac{m^2}{r}\right) \right]
\end{equation}
(an explicit realization with equal blocks $\ell_j = L = N/m$ is analyzed in the next section). For $r \gg m^2$, the subleading correction $\mathcal{O}(m^2/r)$ is suppressed and $V_m(r)$ scales as $(2^m/m!)\, r^m$. Substituting this discrete $L_1$-ball volume for the continuous Euclidean volume in the derivation of Eq.~\eqref{pettis_formula} simply replaces the geometric constant $C_{\rm geom}$ by its discrete counterpart $C_{\rm geom}^{(L_1)}$ and identifies the continuous dimension $d$ with the block count $m$. In practical fits over a finite window, this yields the nearest-neighbor power law of Eq.~\eqref{power-law} with $\alpha = d_{\mathrm{eff}}^{-1}$, where the measured effective dimension $d_{\mathrm{eff}}$ recovers the underlying structural dimension ($d_{\mathrm{eff}} \approx m$) in the ideal regime, while naturally taking non-integer values when blocks have unequal sizes and saturate at different distance scales.

Additional important aspects of this continuous-to-discrete mapping concern the outcome probabilities and the deduplication procedure. In discrete quantum measurements, the unknown continuous probability density $p({\bf x}_i)$ is replaced by the unknown Born probability $P({\bf x}_i) = |\langle {\bf x}_i | \Psi \rangle|^2$ of an individual basis state ${\bf x}_i$, which is likewise independent of the subsample size $n = N_b$ but varies across states in raw measurement records, potentially violating local uniformity and producing duplicate outcomes ($r_1 = 0$). By filtering out repetitions and drawing random subsamples of \textit{unique} configurations ($n = N_b$) uniformly from the resulting support pool [Eq.~\eqref{Hamdist}], every resolved configuration is assigned an equal effective weight within the sampled support. This preprocessing simultaneously establishes the lower bound $r_1 \geq 1$ and a uniform sampling distribution across the support, directly satisfying Pettis's assumption of locally constant density within the first-neighbor ball.

Consequently, although the selection of configurations entering the unique-outcome set partially reflects the underlying Born probabilities at a given measurement budget, uniform subsampling over this set assigns equal effective weights within the resolved support. Indeed, in a discrete state space, empirical frequency weights of individual bitstrings only become statistically meaningful in the oversampled regime ($N_b \gtrsim V_{\mathrm{eff}}$), where nearest-neighbor distances saturate at the discrete lattice spacing ($\bar{r}_1 \sim 1\text{--}2$) and power-law scaling breaks down. Within the undersampled scaling window ($N_b \ll V_{\mathrm{eff}}$), $d_{\mathrm{eff}} = \alpha^{-1}$ therefore characterizes the effective support dimension of the resolved high-probability region rather than the empirical frequency distribution of individual outcomes.

Furthermore, characterizing the sampled support by an effective typical-set volume $V_{\mathrm{eff}}$---defined through the sample average in Eq.~\eqref{pettis_formula} via $(V_{\mathrm{eff}})^{1/d_{\mathrm{eff}}} \equiv \frac{1}{N_b} \sum_{i=1}^{N_b} [p_{\mathrm{eff}}(\mathbf{x}_i)]^{-1/d_{\mathrm{eff}}}$ (with $n = N_b$ and $d \to d_{\mathrm{eff}}$)---yields $\bar{r}_1(N_b) = A N_b^{-\alpha}$ with
\begin{equation}
A = C_{\rm geom}^{(L_1)}\, (V_{\mathrm{eff}})^{1/d_{\mathrm{eff}}}.
\end{equation}
Even when the subsample size is orders of magnitude smaller than the state space ($N_b \ll V_{\mathrm{eff}}$), the average spacing between randomly scattered samples encodes this global scale, so $A$ contains information about the effective accessible state count $V_{\mathrm{eff}}$, which coincides with the full combinatorial support volume $V_{\mathrm{supp}} = |\mathcal{M}|$ for uniform states such as Dicke states, and represents the effective typical-set volume resolved at a given measurement budget for inhomogeneous states.

 Generally, $d_{\rm eff}$ serves as a basis- and scale-specific effective dimensionality descriptor whose applicability is bounded by discrete lattice saturation at small distances and finite-sample sparsity in high-dimensional subspaces, as quantified in the next section. At the same time, the sensitivity of both $A$ and $\alpha$ to the particular quantum state observed in our numerical experiments reflects dimensional compression of the sampled data distribution. This distinguishes our scaling relation from the uncorrelated high-dimensional limits described by Beyer et al.~\cite{Beyer1999}, indicating that the observed power law is a useful descriptor of quantum state measurements. Below, we examine the intrinsic dimension interpretation through concrete numerical examples, provide a deeper analysis of the measurement data for the quantum state families whose representatives are shown in Fig.~\ref{fig1}, and demonstrate possible implementations of our approach.

\section{Estimating the intrinsic dimension}
\begin{figure}[!t]
    \includegraphics[width=\linewidth]{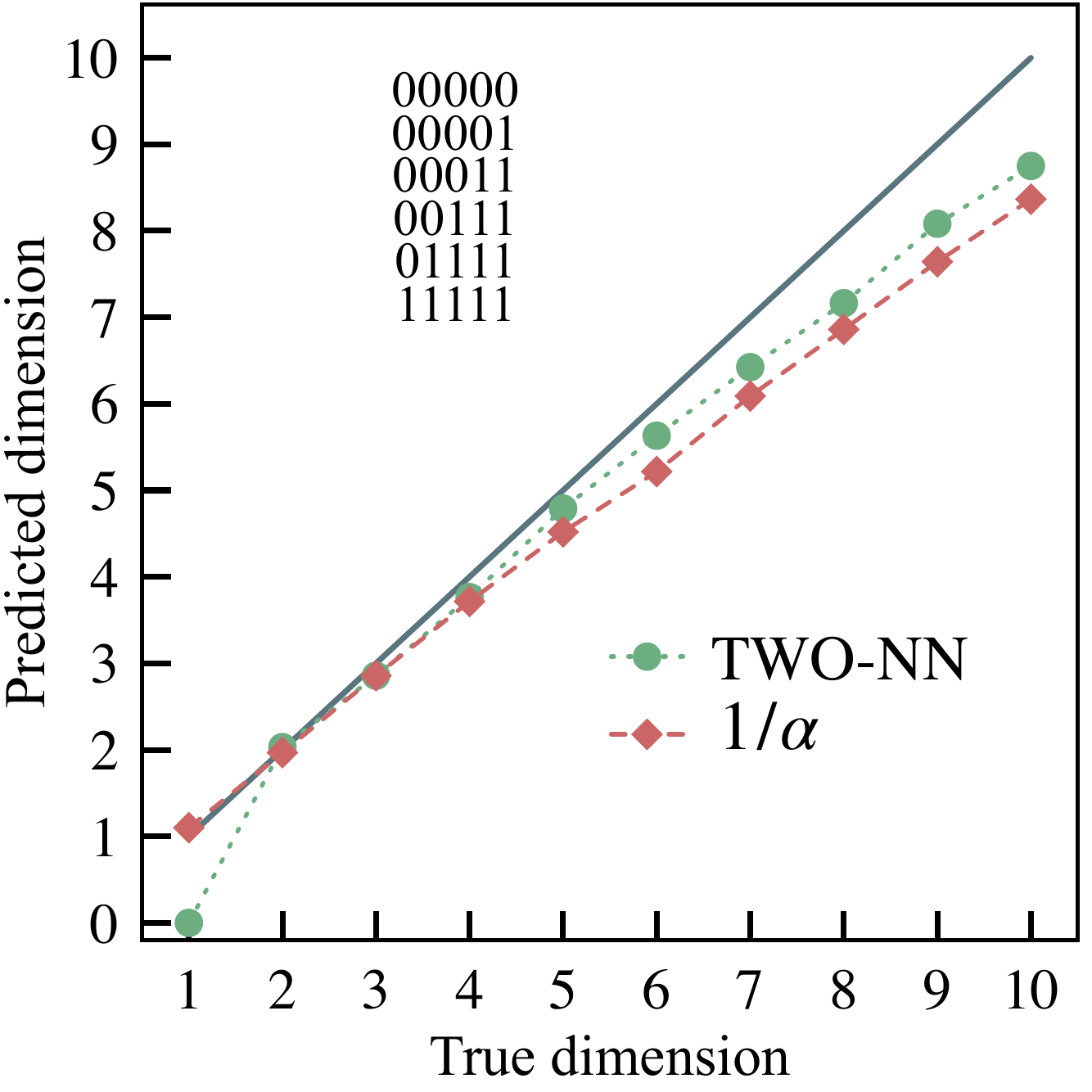}
    \caption{Correlation between the true intrinsic dimension and its estimates obtained by using 2-NN method \cite{TWO-NN} and the proposed power-law approach ($d_{\rm eff} = \alpha^{-1}$). The solid line denotes the ideal scaling baseline. Inset shows the full set of the samples that can be generated with the 5-bit toy model and corresponds to the $m=1$ case.}
    \label{Int_dim}
\end{figure}

To systematically validate the direct relation $d_{\text{eff}} = \alpha^{-1}$ revealed in the previous section, we perform controlled numerical benchmarks on synthetic datasets embedded within a high-dimensional hypercube $\{0, 1\}^N$. By partitioning the available bits, with total system size $N$ ranging slightly from $500$ to $504$ bits to ensure exact divisibility, into $m$ independent and sequentially filled blocks (inset of Fig.\ref{Int_dim}), each generated configuration acts as a coordinate vector on an $m$-dimensional manifold. Since each independent axis of length $N/m$ allows for exactly $(N/m + 1)$ sequential binary configurations, the total number of unique accessible states 
$N_{\text{states}}$ across the entire manifold is given by 
$N_{\text{states}} = \left( \frac{N}{m} + 1 \right)^m$. For instance, 
in the case of $N=500$ and $m=10$, this restricts the valid configuration space to $51^{10} \approx 1.19 \times 10^{17}$ states, which is fundamentally smaller than the hypercube volume of $2^{500}$. We systematically vary the predefined structural dimension $m$ from $1$ to $10$ and for each $m$ we randomly generate 10000 samples. The selection of unique outcomes gives $N_b^{\rm max}$ = 500, 9243, 9986 and 10000 for $m$ = 1, 2, 3 and $m \ge 4$, respectively. Accordingly, to ensure numerical stability, the linear fitting windows for $N_b$ were selected as $[100, 400]$ for $m = 1$ and $[100, 9000]$ for $m \ge 2$.

If our power-law exponent $\alpha$ serves as a true indicator of the underlying manifold dimension, the extracted dimension $d_{\text{eff}} = \alpha^{-1}$ must closely follow the predefined values of $m$. Fig.~\ref{Int_dim} demonstrates that this is indeed the case for $m<6$, where a good agreement between $m$ and $1/\alpha$ is observed. For data of larger dimensions, the deviation of the $1/\alpha$ value from the true intrinsic dimension increases. This behavior can be attributed to the exponential growth of the available state space with increasing $m$, while the number of generated samples remains constant at $N_b^{\rm max} = 10000$. As the data becomes increasingly sparse within the larger multi-dimensional volume, the scaling of the nearest-neighbor Hamming distance experiences finite-size limitations well-documented in manifold learning literature \cite{Beyer1999, Levina2004}. Nevertheless, the monotonic trend across the entire $m=1\dots10$ range confirms that the scaling exponent successfully captures the internal structural geometry of the data. 

To provide an independent validation of our findings, Fig.~\ref{Int_dim} also presents the dimensional estimations obtained via the conventional two-nearest-neighbor (TWO-NN) approach \cite{TWO-NN}, which performs well under the assumption 
that the data density is locally constant. The detailed description of this method is provided in Appendix~B. For the TWO-NN analysis, the intrinsic dimension is evaluated using the entire dataset of unique configurations. For $m \ge 2$, the resulting dependencies of the intrinsic dimension on the number of unique samples saturate (see Appendix~B), demonstrating that the obtained estimation is robust. However, the TWO-NN method fails for $m=1$ because the distance to the first nearest neighbor and that to the second nearest neighbor are strictly equal for every sample due to the discrete nature of the lattice. Remarkably, for all higher dimensions ($m \ge 2$), the TWO-NN curve in Fig.~\ref{Int_dim} demonstrates a close quantitative agreement with our $1/\alpha$ scaling framework across the entire tested dimension range.

It is worth mentioning that there is a conceptual difference between the two methodologies for estimating the intrinsic dimension. Our approach operates by fixing the nearest neighbor consideration while systematically expanding the dataset size $N_b$, which dynamically compresses the radius of the first Hamming sphere. At the same time, the TWO-NN method relies on multiple fixed spatial scales within a single static dataset by evaluating the ratio of the first and second nearest-neighbor distances ($r_2/r_1$) \cite{TWO-NN}. 

These findings suggest a much broader application of our nearest-neighbor-based approach for estimating intrinsic dimensions. As an example, in Appendix C we discuss the results obtained for real-world datasets.    

Taken together, our theoretical derivation and synthetic benchmarks define the scope of the $d_{\rm eff} = \alpha^{-1}$ interpretation. Specifically, mapping the continuum nearest-neighbor formalism \cite{Pettis1979} onto unique bitstrings in $\{0, 1\}^N$ assumes a locally uniform sampling density across the discrete manifold within the selected linear fitting window $N_b \in [N_b^{\rm min}, N_b^{\rm max}]$. While $1/\alpha$ quantitatively reproduces the true manifold dimension in low-dimensional regimes ($m < 6$ for $N_{b}^{\rm max} = 10^4$), finite-sample sparsity in higher-dimensional subspaces leads to systematic underestimation. Consequently, $d_{\rm eff}$ should be interpreted as a basis-dependent, scale-specific effective dimensionality descriptor rather than the exact dimension of the underlying manifold.

\section{Distinguishing quantum states}

\begin{figure}[!b]
    \includegraphics[width=\linewidth]{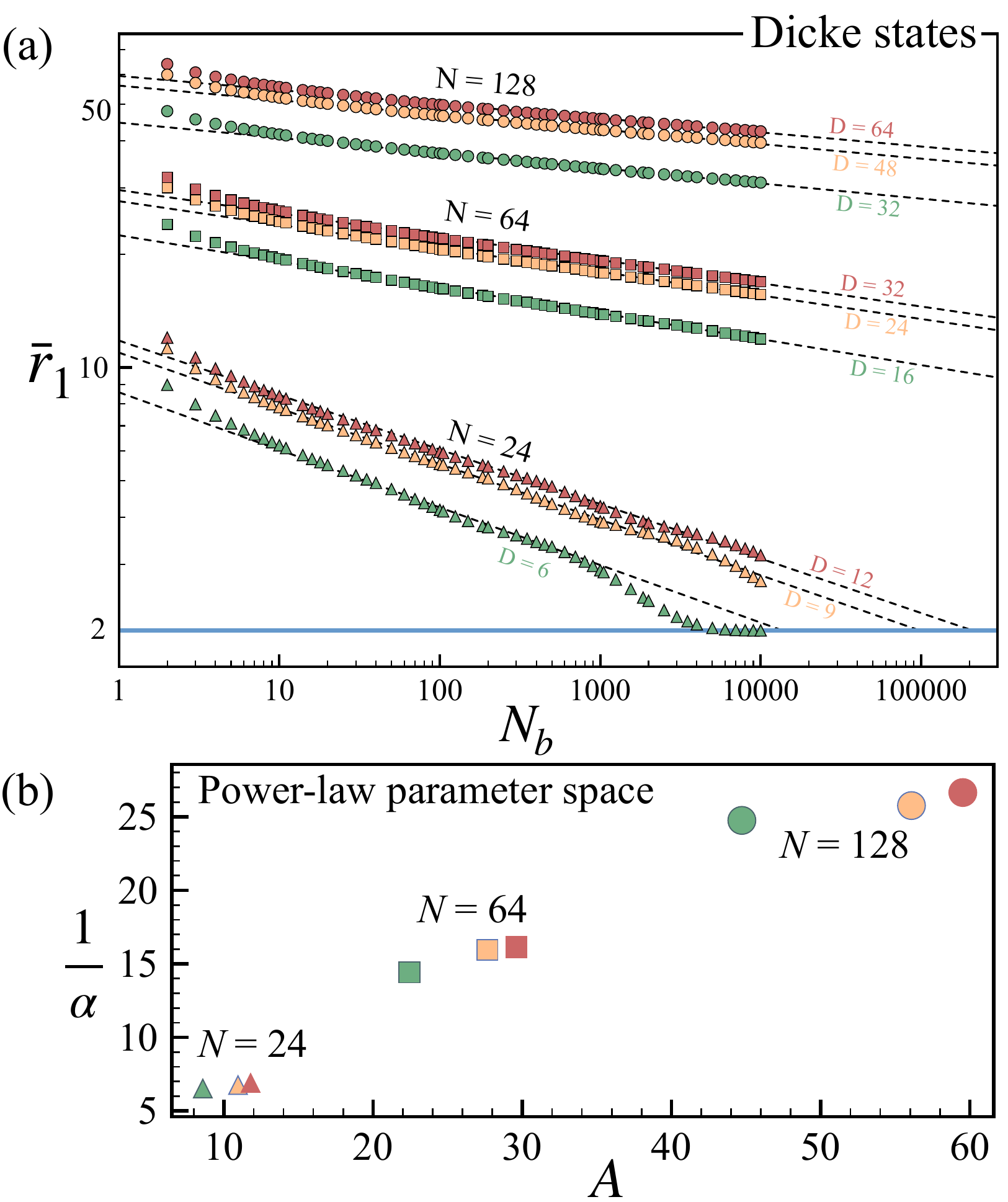}
    \caption{(a) Expected minimal distance function, Eq.~\eqref{Hamdist},  calculated for Dicke states of 24-, 64- and 128-qubit systems in the $\sigma^z$ basis.  Black dashed lines denote the fit of the sampling data with Eq.~\eqref{power-law}. (b) Map distinguishing the different Dicke states with respect to the power-law parameters $A$ and $\alpha$. The error bars with respect to the averaging over non-overlapping datasets of $N_{b}^{\rm max}=10^4$ unique outcomes do not exceed the symbol size.
    }
    \label{Dicke}
\end{figure}

In this section we explain the details of our approach by the example of the Dicke states \cite{Dicke} and then on this basis we will discuss the idea on descriptors of a quantum state. The Dicke wave functions attract considerable attention in quantum computing due to storage \cite{Dicke_storage}, networking \cite{Dicke_network}, sensing \cite{Dicke_sensing}, coding \cite{Dicke_code} and other applications. These permutationally-invariant quantum states are generally defined as 
\begin{eqnarray}
\Ket{\Psi^D_N} = \frac{1}{\sqrt{\binom{N}{D}}} \sum_{j} \mathcal {P}_{j}(\Ket{0}^{\otimes N-D} \otimes \Ket{1}^{\otimes D}).
\label{Dicke_wf}
\end{eqnarray}
Here $N$ is the number of qubits, $D$ is the parameter that controls the complexity of $\Ket{\Psi^D_N}$ through the number of ``1''s in the contributing basis functions and the sum goes over all possible permutations of qubits, denoted with $\mathcal{P}_j$. For the given $N$ the state with $D = N/2$ is characterized by the largest entanglement entropy \cite{dissimilarity}.

Fig.~\ref{Dicke}(a) presents the $\bar{r}_1(N_b)$ functions calculated for the Dicke states of $24$, $64$, and $128$ qubits with different numbers of excitations, specifically $D = N/2$, $3N/8$, and $N/4$. By choosing these specific wavefunctions, we can probe different degrees of data availability when characterizing quantum states.  For instance, for the $24$-qubit Dicke states, a reasonable number of measurements enables the generation of unique bitstrings that almost fully cover the entire available excitation subspace, which is clearly not the case for the $128$-qubit Dicke states. This representative example of the state $\ket{\Psi_{24}^6}$, which possesses the smallest subspace dimension of $134596$, provides a complete picture of the functional patterns of $\bar{r}_1$ at varying sample sizes $N_b$.  The initial segment of the function corresponding to $N_b \in [10, 500]$ is well approximated by the power law $\bar{r}_1(N_b) = 8.59 N_b^{-0.15}$. For $N_b > 1000$, the expected minimal Hamming distance is characterized by a much faster decay and eventually saturates at the absolute minimal distance of $2$ (indicated by the horizontal blue line in Fig.~\ref{Dicke}(a)) when the sample size reaches $N_b = 5000$.

Importantly, the larger the quantum system the more accurate approximation with the power law of the Hamming distance data. In the case of $N= 64$ and 128 we use the range $N_b \in [50, 1000]$ for fitting with Eq.~\eqref{power-law}. From Fig.~\ref{Dicke}(b), which provides the power-law parameters for the Dicke states, one can conclude that the analyzed wavefunctions can be successfully discriminated by means of both scaling parameters. For the $1/\alpha$ 
quantity, which estimates the effective intrinsic dimension of the measurement datasets, the variation of the values obtained for the Dicke states of the same system size $N$ is less pronounced than that observed for the coefficient $A$. To assess the robustness of the obtained power-law parameters with respect to the given set of measurements we divide the whole data for each system to non-overlapping datasets of $N_{b}^{\rm max}=10^4$ unique outcomes. The obtained error bars for both $A$ and $\alpha$ do not exceed the symbol sizes in Fig.~\ref{Dicke}(b). Our results show that the growth of this inverse scaling exponent with the number of excitations at each $N$ closely aligns with the trend of the multipartite entanglement entropy established for this family of wavefunctions \cite{dissimilarity}.

This monotonic increase of the coefficient $A$ with system size $N$ and excitation number $D$ aligns with the theoretical scaling $A = C_{\rm geom}^{(L_1)}\, (V_{\mathrm{supp}})^{1/d_{\mathrm{eff}}}$ derived in the Method section. For a Dicke state $\ket{\Psi_{N}^{D}}$, the accessible support consists of all bitstrings with fixed Hamming weight $D$, giving an exact uniform support volume $V_{\mathrm{eff}} = V_{\mathrm{supp}} = \binom{N}{D}$. Using the fitted effective dimensions $d_{\mathrm{eff}} = \alpha^{-1}$ from Fig.~\ref{Dicke}, the characteristic linear scale $(V_{\mathrm{supp}})^{1/d_{\mathrm{eff}}}$ evaluates to $\binom{64}{32}^{1/16.13} \approx 13.56$ for $\ket{\Psi_{64}^{32}}$ ($A = 29.61$) and $\binom{128}{64}^{1/26.64} \approx 25.30$ for $\ket{\Psi_{128}^{64}}$ ($A = 59.53$). Across these states, the corresponding effective geometric factor $C_{\rm geom}^{(L_1)} = A / (V_{\mathrm{supp}})^{1/d_{\mathrm{eff}}}$ remains of order two and varies only weakly (from $2.20$ to $2.34$) despite the vast growth in $V_{\mathrm{supp}}$.

Now we are in position to discuss the application of our approach to efficient state discrimination. Traditionally, such a problem is mainly solved through quantum state tomography \cite{tomography} which is aimed at reconstructing density matrix on a classical computer and becomes unfeasible in the case of the large-scale quantum systems. This motivates developing alternative procedures that are based on descriptors derived from single-qubit measurements across different bases. In this respect, a prominent example is the dissimilarity measure \cite{dissimilarity} which quantifies the variety of pattern scales present in the bit-string arrays. 

Following this idea we propose to use the set of Hamming-distance-based quantities, $\{ A^{[b]}, \alpha^{[b]} \}$ calculated in different bases $b = 1 .. N_{B}$ as a distinct geometric descriptor of a quantum state. To demonstrate its practical utility, we compare the half-filled Dicke state, $\ket{\Psi_{24}^{12}}$, against a trivial uniform product state, $\ket{\Psi_{\rm {u}}} = [(\ket{0}+\ket{1})/\sqrt{2}]^{\otimes 24}$, and an entangled Haar-random state, $\ket{\Psi_{\rm {Haar}}}$, in a $24$-qubit system. 
Being strongly delocalized across the computational $\sigma^z$ basis, 
these states present a notorious challenge for conventional neural-network quantum state tomography \cite{Torlai, ourRBM} and certification protocols \cite{Sycamore}.

\begin{figure}[!t]
    \includegraphics[width=\linewidth]{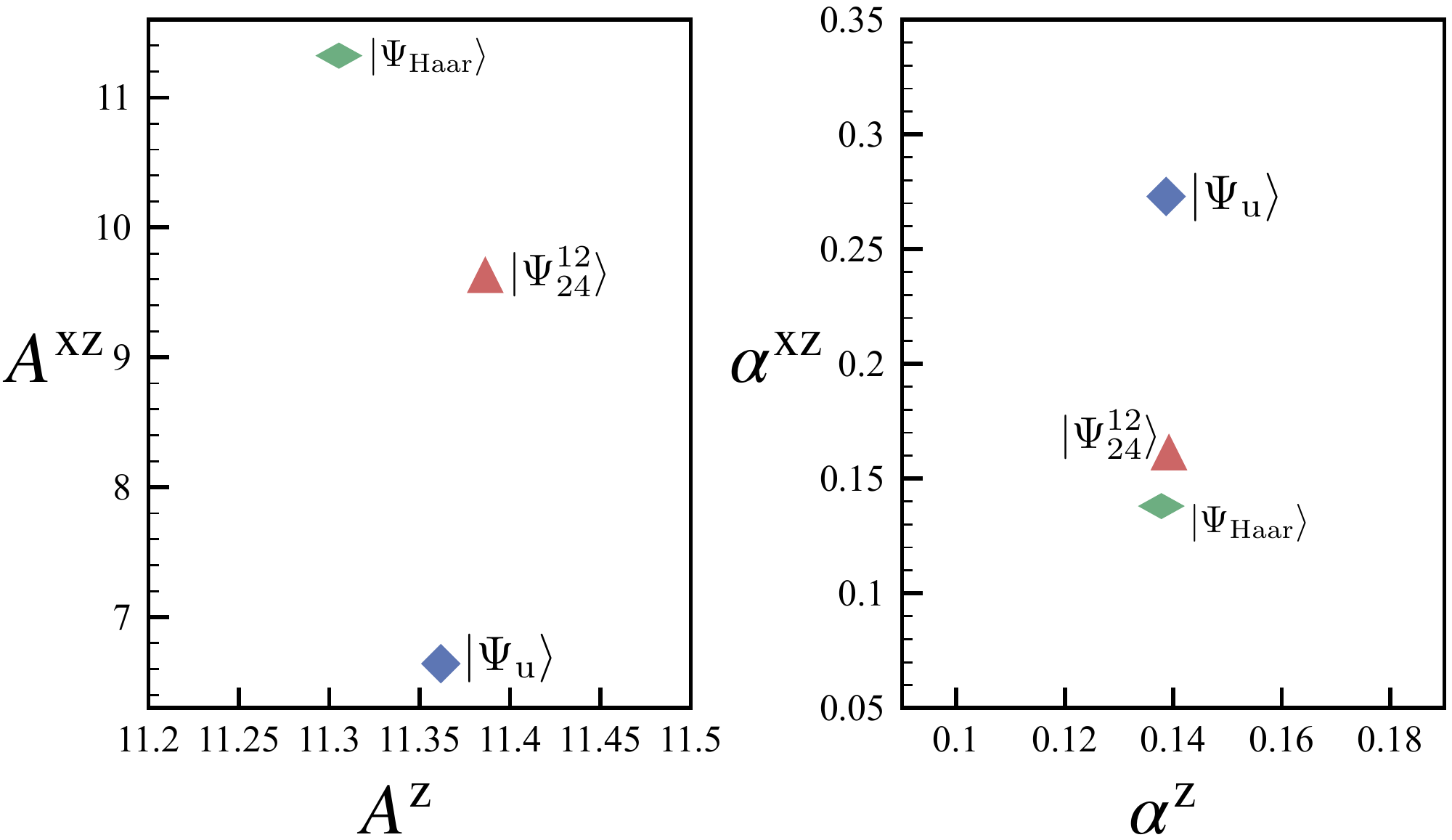}
    \caption{Power-law parameters calculated for strongly delocalized quantum states of 24 qubits. $\ket{\Psi^{12}_{24}}$, $\ket{\Psi_{\rm u}}$ and  $\ket{\Psi_{\rm Haar}}$ denote Dicke, uniform and Haar-random wave functions, respectively.}
    \label{bases}
\end{figure}

Fig.~\ref{bases} illustrates the power-law parameters estimated from measurements in two bases: the standard computational $\sigma^z$ basis, and a hybrid $\sigma^x - \sigma^z$ basis combining $\sigma^x$ and $\sigma^z$ projections for odd and even qubits, respectively. For fitting with Eq.~\eqref{power-law} we use the range $N_b \in [50, 1000]$. Remarkably, while the considered wavefunctions are practically indistinguishable when analyzed solely via $\sigma^z$ basis measurements, switching to the hybrid $\sigma^x - \sigma^z$ basis enables their clear discrimination.
It is worth mentioning that the power-law parameters for the Haar-random states remain nearly invariant with respect to the choice of the measurement basis, yielding $A^{\rm z}_{\rm Haar} \approx A^{\rm xz}_{\rm Haar}$ and $\alpha^{\rm z}_{\rm Haar} \approx \alpha^{\rm xz}_{\rm Haar}$. Additionally, the extracted exponents $\alpha^{\rm xz}$ correspond to effective intrinsic dimensions of $d_{\rm eff} = 7.20$ and $3.66$ for the Haar-random and uniform states, respectively. This quantitative differentiation perfectly aligns with alternative complexity measures characterizing these multi-body wavefunctions \cite{dissimilarity}.

\section{Phase boundaries of $J_1-J_2$ model solutions} 
The possibility of distinguishing quantum states with a limited number of measurements paves the way to a solution of phase classification problem which holds a significant importance in condensed matter physics. To demonstrate this we use a nontrivial example of the two-dimensional $J_1-J_2$ model with antiferromagnetic interactions between nearest ($J_1$) and next-nearest ($J_2$) neighbours on the $\sqrt{N} \times \sqrt{N}$ square lattice,
\begin{eqnarray}
\label{Heis_Ham}
H_{J_1 - J_2} = J_1 \sum_{\braket{i,j}} {\bf S}_{i} {\bf S}_{j} +  J_2 \sum_{\braket{\braket{i,j}}} {\bf S}_{i} {\bf S}_{j}.
\end{eqnarray} 
Depending on the value of $J_2$ the model is characterized by phases of different complexity. For instance, the quantum analogs of the classical N\'eel and stripe phases were found for $J_2 \in [0, 0.45]$ and $J_{2} \in [0.61, 1]$, respectively. Despite considerable efforts, there is still no consensus on the ground states of the $J_1 - J_2$ model in the intermediate regime $J_2 \in (0.45, 0.61)$, which represents a special difficulty for numerical simulations \cite{Imada}. To find the ground states of the $J_1-J_2$ model on the square lattice with periodic boundary conditions and $N = 36, 64$ and 100 spins $\frac{1}{2}$, we use the exact diagonalization (ED) \cite{spinED} and variational neural-network quantum state approach utilizing a Vision Transformer network (ViT) for wave function representation \cite{ViT1, ViT2} as realized in the NetKet package \cite{Netket2019, Netket2022}. The details on the training and sampling procedures are provided in Appendix D. 

\begin{figure}[!b]
    \includegraphics[width=\linewidth]{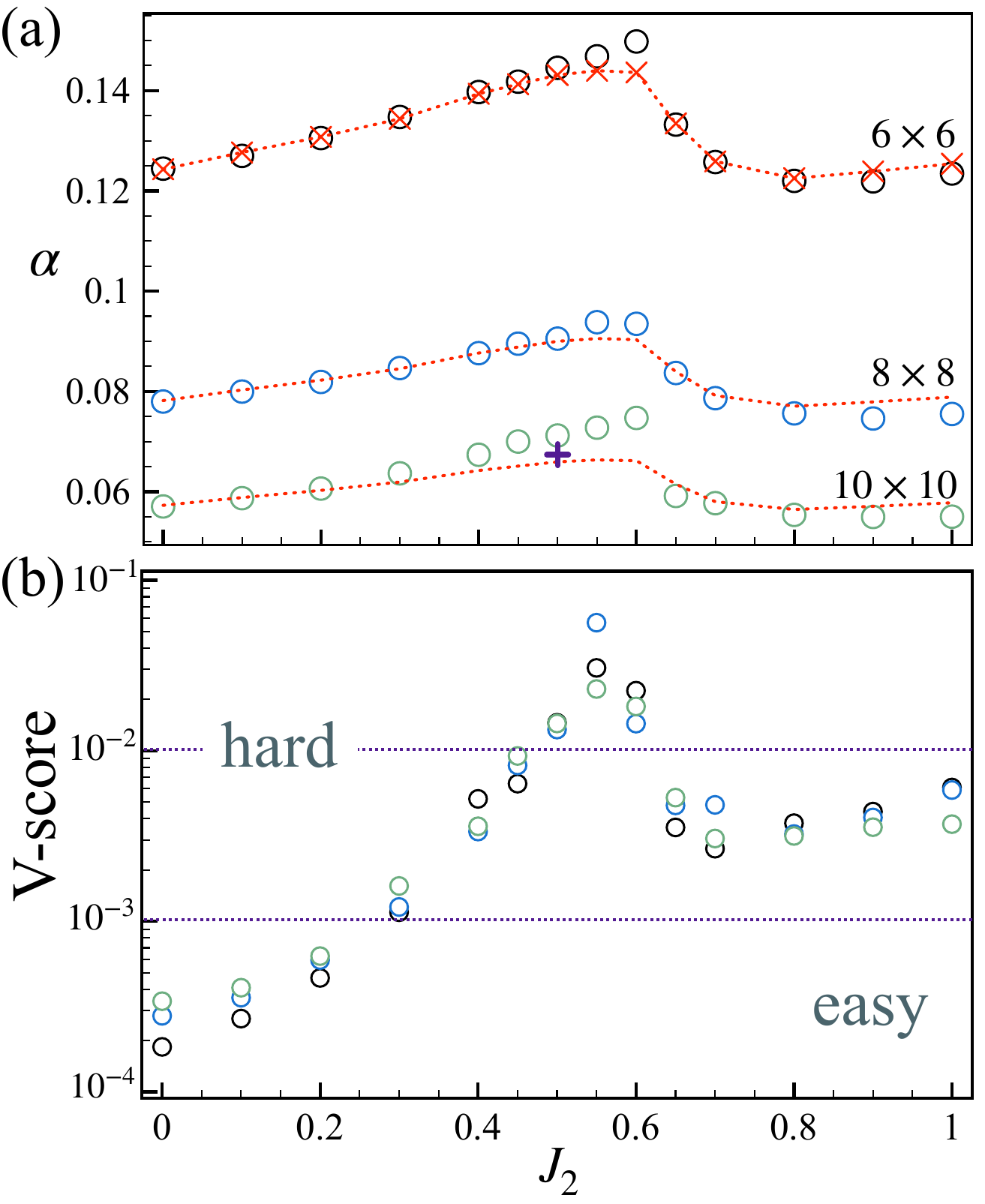}
    \caption{(a) Power-law parameter, $\alpha$ calculated for exact ED and approximated ViT ground states of the antiferromagnetic $J_1-J_2$ model. Black, blue and green circles correspond to the ViT results obtained for $6 \times 6$, $8 \times 8$ and $10 \times 10$ supercells, respectively. Red crosses stand for ED solutions of the $6 \times 6$ system. Red dotted lines denote ED data rescaled as described in the text. The violet plus sign corresponds to the estimate based on pre-trained ViT data from \cite{DeepViT}. The errorbars with respect to the averaging over non-overlapping datasets of $N_{b}^{\rm max}=2\cdot10^4$ unique outcomes are smaller than the symbol size. (b) V-score metric calculated  for the ViT solutions with Eq.~\eqref{V-score}. }
    \label{Heis}
\end{figure}

ED results obtained for the $J_1 - J_2$ model on the $6 \times 6$ supercell (the red crosses in Fig.~\ref{Heis}) indicate that the power $\alpha$ is sensitive to critical regions of the model phase diagram. In particular, it exhibits a linear behavior for small $J_2 \in [0, 0.3]$ and a smooth maximum spanning the known critical area ($J_2 \in [0.5, 0.6]$), with the largest values attained at $J_2 = 0.6$. The kink observed at this point aligns with the transition region reported in prior studies between the intermediate and stripe phases \cite{Imada}, serving as a complementary geometric signature of changing ground-state correlations. The demonstrated connection between the exponent $\alpha$ and the intrinsic dimension of the sampling datasets supports the interpretation that this behavior of the exponent $\alpha$ reflects the critical behavior of the system in question. Previous works \cite{Dalmonte1, Dalmonte3, Dalmonte4} have revealed that the intrinsic dimension features extrema at the point of equilibrium and non-equilibrium transitions.

These ED results provide a solid basis for analyzing the approximate solutions obtained with the variational ViT quantum state approach. From Fig.\ref{Heis} we observe an excellent agreement between ED and ViT values of the power-law parameter for $J_{2} \in [0, 0.4]$ (the N\'eel phase). Expectedly, the largest deviations from the exact solution takes place at the critical $J_2 = 0.55$ and $J_2 = 0.6$, which agrees with largest deviations in ground state energies obtained for these points (Appendix D).  

Increasing the system size leads to a suppression of the absolute values of the $\alpha (J_2)$ function. Nevertheless, the qualitative profiles of the function are robust with the maximum near $J_2 = 0.6$. To demonstrate trends more explicitly, we perform a scaling of the ED dependencies obtained for the $6 \times 6$ supercell (red dotted lines in Fig.~\ref{Heis}). For each system size, the scaling factors are defined to match the ViT data at $J_{2} = 0$. This represents the simplest, non-frustrated regime for simulations, where reliable quantum Monte Carlo (QMC) results \cite{Sandvik_2026} are available. The accuracy of our ViT results at $J_{2} = 0$ can be demonstrated by calculating the relative errors of the ground-state energies with respect to the QMC ones (Appendix D). Fig.~\ref{Heis} shows that the ViT power-law parameters (green and blue circles) closely follow the scaled $6 \times 6$ dependencies (red dotted lines) for $J_{2} \in (0, 0.3]$. Such agreement is consistent with the expectation that the ViT provides an accurate approximation of the unknown ground states in this parameter regime. In turn, outside this parameter range ($J_{2} > 0.3$) we observe deviations of the $\alpha$ values from the scaled curves, which become especially pronounced for the $10 \times 10$ system. Importantly, at $J_2 = 0.5$ a pre-trained ViT (pViT) with 434760 parameters \cite{DeepViT} is available for a $10 \times 10$ system; this model yields a lower ground-state energy, $E_{\rm pViT} = -0.497508$, than that obtained in our ViT calculations with 155620 parameters, $E_{\rm ViT} = -0.496783$. Moreover, the deviation of the scaling exponent $\alpha$ from the rescaled $6 \times 6$ curve (indicated by the purple plus marker in Fig.~\ref{Heis}) becomes noticeably smaller in comparison with our standard ViT solution. Specifically, the values shift from $\alpha_{\rm ViT} = 0.071\pm0.001$ to $\alpha_{\rm pViT} = 0.067\pm0.001$ toward the rescaled baseline of $\alpha^{6\times 6}_{\rm rescaled} = 0.066\pm0.001$. The errors are given with respect to the averaging over non-overlapping datasets of $N_{b}^{\rm max}=2\cdot10^4$ unique outcomes. Within this benchmark setting, the concurrent improvement in variational energy and closer agreement with the rescaled ED curve provide a useful model-specific consistency check, suggesting that our geometric metric can complement standard energy-based indicators when tracking optimization progress within a fixed model family.

Traditionally, the system with $J_2 = 0.5$ is considered to be the most frustrated one and is used as a benchmark point for different variational eigensolvers \cite{DeepViT, SSE_Heyl}. According to our results, the largest deviations from the $6 \times 6$ baseline are observed at $J_2 = 0.55$ and 0.6, which suggests treating the solutions at these points to be particularly challenging for ViT-based neural quantum state approaches. To confirm these conclusions we have calculated the dimensionless V-score metric \cite{V-score} that combines the variational energy, $E$ and energy variance, ${\rm Var} E = \braket{\hat H^2} - \braket{\hat H}^2$ estimated for an $N$-qubit system. For the $J_1 - J_2$ spin model such a metric is given by
\begin{eqnarray}
\label{V-score}
{\rm V-score} = \frac{N {\rm Var} E}{E^2},
\end{eqnarray}
and is a reliable estimator for the order of magnitude of the energy relative error \cite{V-score}. By using V-score, the case of the $J_1 - J_2$ square lattice Hamiltonian was identified as a hard problem for variational algorithms where one cannot expect an energy accuracy better than one or two significant digits. Our ViT calculations presented in Fig.\ref{Heis} (b) confirm this for the intermediate values of $J_2$. In the regime of zero ($J_2=0$) or weak ($J_2 \in (0, 0.2]$) frustration, the V-score values are of the order of $10^{-4}$, which agrees with the previous results \cite{V-score}. For all the considered systems the V-score functions feature maxima at $J_{2} = 0.55$ on the order of $10^{-2}$, which correlates with the behaviour of the $A(J_2)$ function (Appendix D). Thus, the power-law parameters not only capture geometric signatures of phase boundaries, but also serve as a complementary model-specific consistency check for variational ground-state approximations. 

\begin{figure}[!b]
    \includegraphics[width=\linewidth]{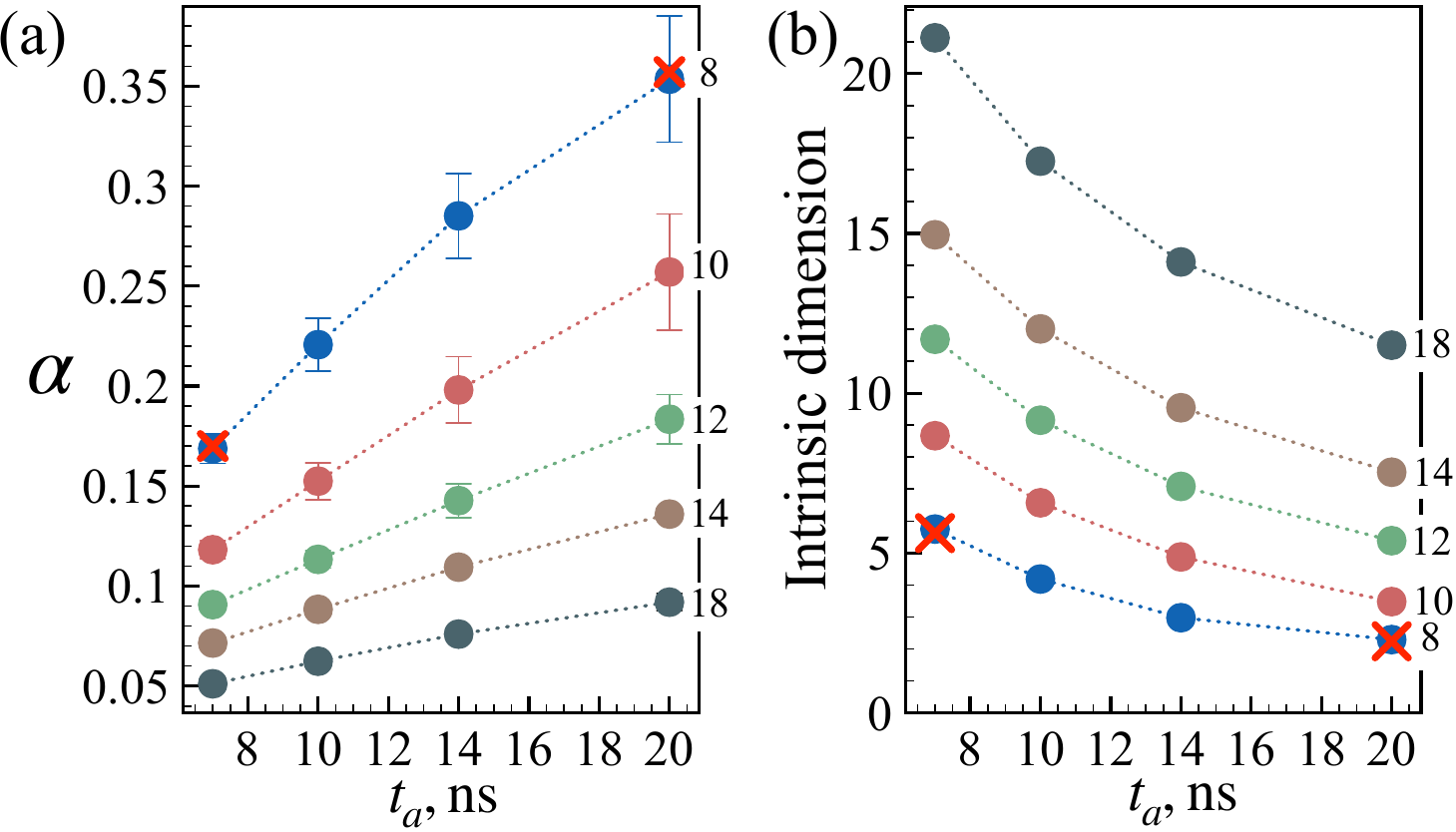}
    \caption{(a) Dependence of the power $\alpha$ on the annealing time calculated for D-Wave sampling data \cite{DWave2025}. Numbers on the right indicate linear system sizes. These results were averaged over 20 independent instances of disorder. (b) TWO-NN intrinsic dimension averaged over 20 independent instances of disorder. Red crosses denote the power $\alpha$ (a) and ID (b) calculated for MPS $8 \times 8$ samples from Ref.\onlinecite{DWave2025}. }
    \label{D-Wave}
\end{figure}

Finally, we clarify the role of the measurement basis in interpreting these $J_1 - J_2$ results. As highlighted in recent wave-function network studies~\cite{Dalmonte3}, intrinsic-dimension descriptors are inherently basis-dependent: a quantum state can appear maximally delocalized and nearly incompressible in an uninformative basis, whereas measuring in a physically informative basis aligned with the dominant correlations compresses the bitstring support and exposes critical behavior (as also illustrated in Fig.~\ref{Heis}). For the isotropic $J_1 - J_2$ Hamiltonian [Eq.~\eqref{Heis_Ham}], the exact finite-size ground states are spin-singlet states with zero total spin ($S_{\mathrm{tot}} = 0$), satisfying $U^{\otimes N}|\Psi_{\mathrm{GS}}\rangle = |\Psi_{\mathrm{GS}}\rangle$ up to a global phase under any uniform single-spin rotation $U \in SU(2)$. Consequently, the projective measurement distribution and the resulting parameters $(A, \alpha)$ are strictly invariant under \textit{global} basis rotations (e.g., from $\sigma^z$ to $\sigma^x$ or $\sigma^y$). Crucially, this stability follows from the global singlet symmetry of the states rather than from basis independence of our Hamming-distance framework: under inhomogeneous (site-dependent) local basis transformations $\bigotimes_{i=1}^N U_i$ or for non-singlet states, the bit-string support can become strongly delocalized, potentially obscuring phase-boundary signatures. Moreover, because our variational ViT wave functions enforce $S^z_{\mathrm{tot}} = 0$ rather than exact total-spin singlet symmetry ($S_{\mathrm{tot}} = 0$), invariance under global basis rotations holds only approximately for these variational states, depending on how accurately the optimization reproduces the underlying singlet correlations.

\section{Intrinsic dimension of the quantum annealing data} 
To demonstrate the versatility of our approach in processing data from different sources, next we analyze the samples obtained in the recent D-Wave quantum annealing experiments \cite{DWave2025}, where a quantum processing unit (QPU) is programmed to generate samples in close agreement with solutions of the Schr\"odinger equation for the Ising model Hamiltonian 
\begin{eqnarray}
\label{D-Wave_Ham}
H (t) = - \Gamma(t/t_a) \sum_{i} \sigma^x_{i} + \mathcal J(t/t_a) \sum_{ij} J_{ij} \sigma^z_{i} \sigma^z_{j}.
\end{eqnarray}
Here $\sigma^{\mu}_{i}$ ($\mu = x,z$) is the Pauli matrix, the couplings $J_{ij}$ are chosen randomly with respect to the different topologies. The time-dependent coefficients $\Gamma(t/t_a)$ and $\mathcal J(t/t_a)$ control the Ising and transverse-field contributions in Eq.~\eqref{D-Wave_Ham} that have opposite schedules within annealing procedure ($\Gamma(0) >> \mathcal J(0)$ at $t=0$ and $\Gamma(1) << \mathcal J(1)$ at the quench time $t=t_a$). 

A system undergoing a quantum phase transition during Schr\"odinger evolution is described by the quantum Kibble-Zurek mechanism \cite{Polkovnikov, Zurek, Dziarmaga}, which predicts the formation of the defects due to the finite quench time, $t_a$. If the Ising part in Eq.~\eqref{D-Wave_Ham} described a spin chain with uniform ferromagnetic  nearest-neighbor interactions, where the ground state corresponds to collinear ordering, the defects in the measurement results could be easily detected and characterized, as demonstrated in experiments with quantum annealers \cite{King2000} and Rydberg atom platforms \cite{Keesling}.  For the spin glasses, Eq.~\eqref{D-Wave_Ham} with random exchange interactions, the situation is more challenging, since the ground-truth states can be found only for small-scale systems (for instance, the $6\times 6$ square lattice). Fortunately, the spin-glass order parameter $\braket{q^2}$ \cite{Edwards_Anderson} allows quantifying the degree of spin-freezing. Its increase toward unity signifies a reduction in the number of defects and a closer approach to the ideal spin-glass configuration. 

In turn, our framework facilitates a better understanding of the configuration space structure and its dynamic evolution with the quench time $t_a$. Specifically, we analyze experimental data \cite{DWave2025} collected for a system with a square lattice topology, as it provides the largest available set of unique bitstrings. The expected minimal Hamming distance functions $\bar{r}_1(N_b)$ calculated for the $18 \times 18$ spin system at different quench times (see Fig.~\ref{fig1}) exhibit a well-defined power-law behavior. The scaling rate can be quantitatively 
evaluated via the exponent $\alpha$ extracted from the power-law fitting of $\bar{r}_1(N_b)$. This power-law scaling holds for smaller models as well (Fig.~\ref{D-Wave}(a)), indicating that a longer quench time $t_{a}$ leads to a stronger localization of the sampled measurement outcomes within a lower-dimensional subset of the configuration space, which is consistent with the suppression of Kibble-Zurek defects and the convergence toward low-energy spin-glass solutions. Here, we use $N_b \in [50, 700]$ and $N_b \in [50, 900]$, depending on the specific number of available unique bitstrings. The extracted exponent is directly linked to the effective intrinsic dimension of the sampling data manifold via $d_{\text{eff}} = \alpha^{-1}$. 
As the quench time $t_a$ increases, the steady growth of $\alpha$ translates into a significant reduction of $d_{\text{eff}}$, capturing how quantum fluctuations compress the delocalized initial state into a compact, low-dimensional subset of target configurations. 

To confirm our conclusions regarding the intrinsic dimension behavior in the D-Wave system, we independently evaluate it using the conventional TWO-NN approach \cite{TWO-NN}. To calculate the $d_{\rm{TWO-NN}}$ for each system size $L$ we use a fixed amount of unique bitstrings for all $t_\alpha$. Namely, $N_b$ = 500, 800 and 900 for $L$ = 8, 10-14 and 18, respectively. The obtained results explicitly highlight the extreme redundancy of the initial $N$-dimensional configuration space representation. For instance, at $t_a = 7$~ns, the estimated intrinsic dimension is of the order of the linear system size,  $d_{\rm TWO-NN} \sim \sqrt{N}$ [Fig.\ref{D-Wave} (b)]. The obtained results show that $d_{\rm{TWO-NN}}$ further decreases with $t_a$, demonstrating that the effective dimensionality of the subspace encompassing the system's wavefunction outcomes becomes systematically smaller during the annealing process. For all considered datasets, the intrinsic dimension extracted via the TWO-NN method is again in close quantitative agreement with the $1/\alpha$ scaling framework.    

The quench dynamics of small-size spin-glass systems can be accurately reproduced with classical approaches such as the matrix product state (MPS) \cite{DWave2025}. We likewise observe excellent agreement between MPS and QPU Hamming-distance quantities characterizing state space [Fig.\ref{D-Wave}] of system with the $8 \times 8$ square lattice. Simulations of larger systems with the same or more complex topologies represent a significant challenge for classical methods. Here, information on state-space properties extracted from experimental data could prove useful for further optimization of classical algorithms. For instance, our results demonstrate a simplification of the wave-function structure with $t_a$, which motivates the search for more effective classical representations of quantum wave functions with fewer parameters similar to the static case \cite{ourRBM, Cirac_Dicke, Clark, Carleo_exact, Duan}.     

\section{Conclusions} 
In conclusion, we have revealed a common geometric feature characterizing quantum measurement data obtained from various sources, which manifests as a power-law behavior of the expected minimal Hamming distance between the unique configurations with respect to the number of samples. This robust scaling behavior enables the extraction of inherent data manifold characteristics, such as the power-law parameters $A$ and $\alpha$, independently of the total sample size. By adapting nearest-neighbor spatial statistics to unique bitstring ensembles, we have linked the scaling exponent $\alpha$ to an effective, basis-dependent intrinsic dimension $d_{\text{eff}} = \alpha^{-1}$ of the sampled data manifold and established its practical validity limits imposed by finite sample sizes and manifold sparsity. 

Within these well-defined regimes, our numerical benchmarks across synthetic manifolds, frustrated spin models, quantum annealing experiments, and classical image datasets show that the empirical parameters $(A, \alpha)$ serve as practical geometric descriptors for state discrimination, variational ansatz benchmarking, and detecting signatures of critical behavior. The introduced Hamming-distance parameters provide direct information on the spatial localization within the configuration space and thus complement standard problem-specific observables, such as magnetic order parameters.

In Appendix~C, we extend our scaling analysis onto the real-world classical MNIST dataset to further verify the structural versatility of our geometric approach.

\section{Acknowledgments}
We would like to thank Oleg Sotnikov for technical assistance. This work was supported by the Russian Science Foundation, Grant No. 26-12-00377, https://rscf.ru/project/26-12-00377/.

\section*{AI Use Disclosure}
The authors used Google Gemini during the revision stage to assist with literature search, English-language editing, and refining the clarity of physical explanations and claims. All calculations, simulations, and data analyses were conducted independently by the authors, who thoroughly verified all revised text and take full responsibility for the content of the work.

\subsection{Appendix A: Sample redundancy and linear fitting windows}

\begin{figure}[!t]
    \includegraphics[width=\linewidth]{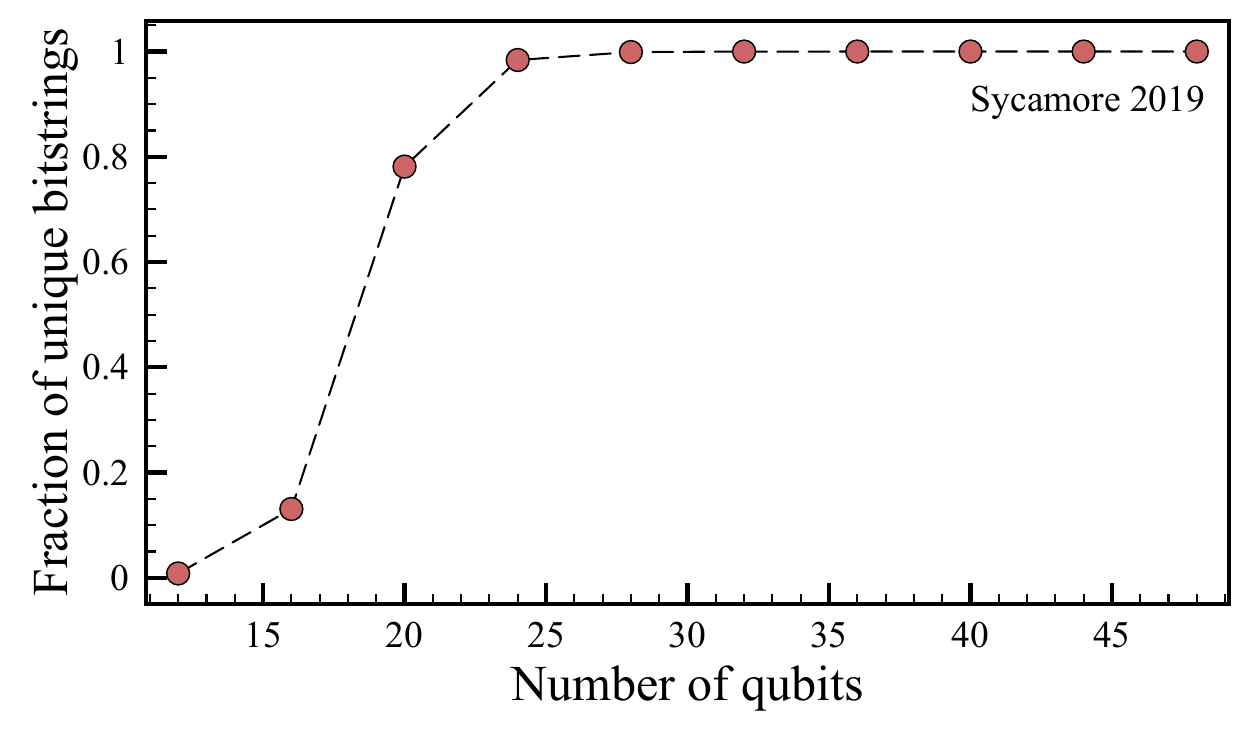}
    \caption{Fraction of unique bitstrings across the datasets obtained with Google Sycamore in Ref.\cite{Sycamore}. For each system size the total number of sampled bitstrings is 500000.}
    \label{repetitions}
\end{figure}

{\it Bit-string repetitions.}
Using unique bitstrings accounts for the typical sampling regime of large-scale, delocalized quantum systems. The occurrence of repeated outcomes depends not solely on the nominal Hilbert-space dimension \(2^{N}\), but on the delocalization of the underlying probability distribution in the measurement basis relative to the total number of measurements. For strongly delocalized states,  such as the Haar-random states sampled on the Google Sycamore processor (\(5 \times 10^5\) measurements, Fig. \ref{repetitions}) -- the probability of observing duplicate bitstrings becomes negligible for \(N \geq 30\), making direct estimation of individual state probabilities infeasible. Conversely, for strongly localized states or data constrained to low-dimensional manifolds (such as our $N = 100$, $m = 1, 2$ synthetic benchmarks discussed below), repeated outcomes still occur frequently even at large $N$, which requires explicitly filtering for unique configurations prior to computing nearest-neighbor distances.

\begin{figure}[!t]
    \includegraphics[width=\linewidth]{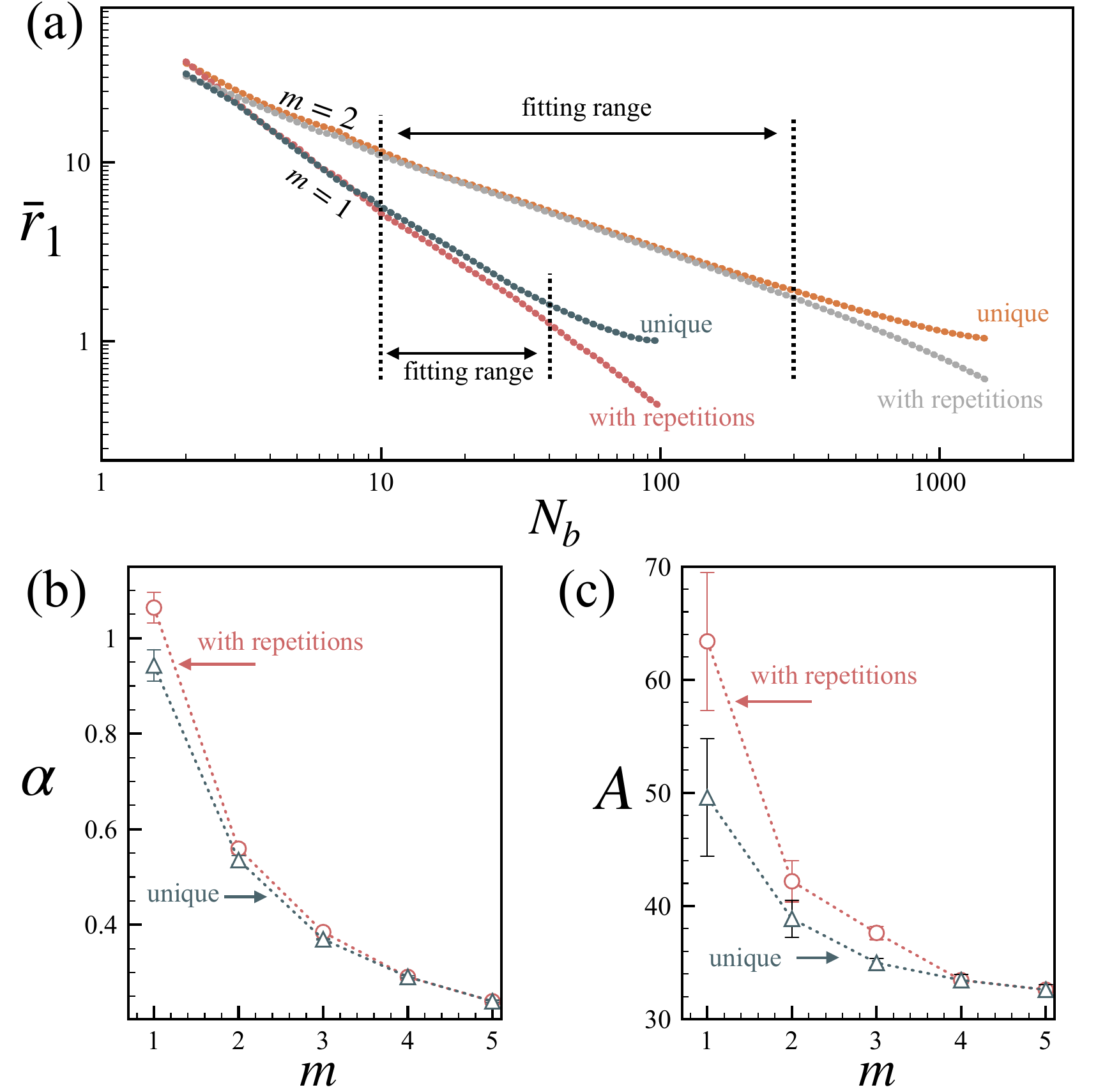}
    \caption{Comparison of power-law results obtained with and without bit-string repetitions for $100$-bit synthetic datasets generated for an $m$-dimensional manifold. (a) Examples of the $\bar r_1 (N_b)$ functions calculated for $m=1$ and $m=2$ datasets. Vertical dotted lines denote the fitting ranges we used. (b)-(c) The power-law parameters estimated over 25 independent runs.} 
    \label{100-bit}
\end{figure}

Retaining repeated bitstrings, if they are present in the measurement dataset, alters the scaling. While the log-log plot remains approximately linear, the straight line exhibits a faster decay, leading to a steeper slope and a higher value of $\alpha$ (as well as a larger extrapolated prefactor $A$). We demonstrate this with Fig.~\ref{100-bit} using the example of $100$-bit synthetic datasets generated for an $m$-dimensional manifold by employing the same approach discussed in the main text. In these simulations, we generate $10000$ bitstrings in total, and the corresponding number of unique outcomes is equal to $100$ ($m=1$), $2437$ ($m=2$), $8755$ ($m=3$), $9862$ ($m=4$), and $9983$ ($m=5$). The repetition-based estimates of $\alpha$ and $A$ are larger than those obtained for unique bitstring datasets, especially for the $m=1$ and $m=2$ cases, where the number of repetitions allows one to approximate the probabilities of individual outcomes. Conversely, for $m=4$ and $m=5$ with almost zero duplication, both procedures yield virtually identical parameters, confirming that the steeper log-log scaling is a general consequence of non-negligible duplicate rates rather than an artifact specific to our synthetic model. For $m=2$, the estimate based on the unique bitstrings ($\alpha^{-1} = 1.87\pm0.03$) provides a slightly better agreement with the predefined value of the intrinsic dimension than that based on the full dataset with repetitions ($\alpha^{-1} = 1.79\pm0.03$).

{\it Selection of fitting ranges.}
To define the fitting range, we first construct the $\bar r_1 (N_{b})$ function in the interval $N_{b} \in [2, N_b^{\text{max}}]$, where $N_b^{\text{max}}$ is the number of available unique bitstrings. For all systems considered in our work, we observe a stable linear part of the $\bar r_1 (N_{b})$ function in the log-log scale starting at a certain threshold $N_b^{\text{min}}$. To ensure the independence of the resulting parameters $A$ and $\alpha$ from the choice of $N_b^{\text{min}}$, we additionally vary it within a local range. The resulting fitting ranges for all the systems under consideration are explicitly specified in the text.

\subsection{Appendix B: TWO-NN method for estimating intrinsic dimension} 
The initial dimension of the feature-vectors describing the system in question is usually excessive and the data can be embedded in lower-dimensional space. The dimensionality of such a subspace is called an Intrinsic Dimension~\cite{Id} and can be associated with the minimal amount of the degrees of freedom and, therefore, with Kolmogorov complexity~\cite{Kolmogorov}. 

\begin{figure}[!b]
    \includegraphics[width=\linewidth]{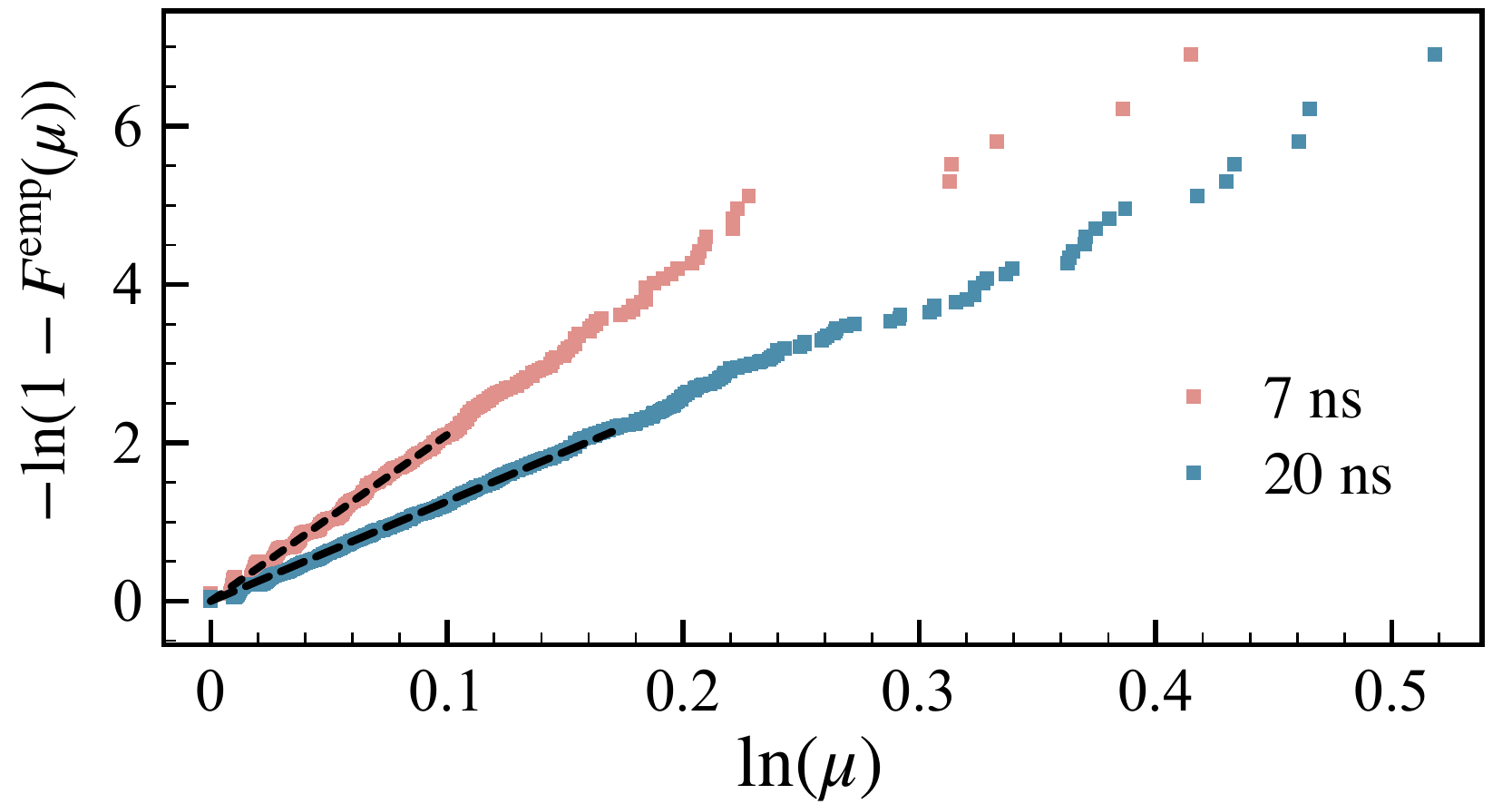}
    \caption{Example of cumulative distributions against $\mu$ obtained for one specific disorder realization for the 18x18 D-Wave data at $t_a = 7$ ns and $t_a = 20$ ns. The corresponding $d_{\rm TWO-NN}$ values defined from fit are equal to 21 and 12.59, respectively. }
    \label{ID_distr}
\end{figure}

In the TWO-NN estimator method~\cite{TWO-NN}, the intrinsic dimension is defined with the following equation
\begin{eqnarray}
\label{Id}
d_{\rm TWO-NN} = - \frac{\ln[1 - F^{\rm emp}(\mu)]}{\ln(\mu)},
\end{eqnarray}
where $F^{\rm emp}(\mu)$ is the empirical cumulative distribution function and $\mu=r_2/r_1$ is calculated for each N-dimensional vector from the dataset, where $r_1$ and $r_2\ge r_1$ are the distances to the nearest and next-nearest neighbours, respectively. In the initial formulation, Euclidean distance between data points is used. However, nothing, in general, prevents from defining any other relevant metrics depending on the particular task. For instance, in the case of quantum measurements data it is common to use Hamming distance as it is a physically relevant measure~\cite{Dalmonte1}. It is important to note, that the TWO-NN algorithm by construction works for datasets of unique samples, since any duplications will immediately lead to the division by zero in $\mu=r_2/r_1$. 

To calculate the $d_{\rm TWO-NN}$ of the dataset including $N_b$ bitstrings one needs to sort $\mu$ values in the ascending order. After doing this, the empirical cumulative distribution function corresponding to $i$th bitstring in the sorted dataset can be defined as $F^{\rm emp}(\mu_i) = \frac{i}{N_b}$. Thus, the resulting points will have coordinates $\{-\ln[1 - F^{\rm emp}(\mu_i)], \ln(\mu_i)|i=1,\dots,N_b\}$ and should be fitted with the straight line passing through the origin. The intrinsic dimension is simply the slope of this line (Fig. \ref{ID_distr}). To make the procedure more robust, we discard the 10\% of the points characterized by highest values of $\mu$ from the fitting as it was proposed in \cite{TWO-NN}.

\begin{figure}[!t]
    \includegraphics[width=\linewidth]{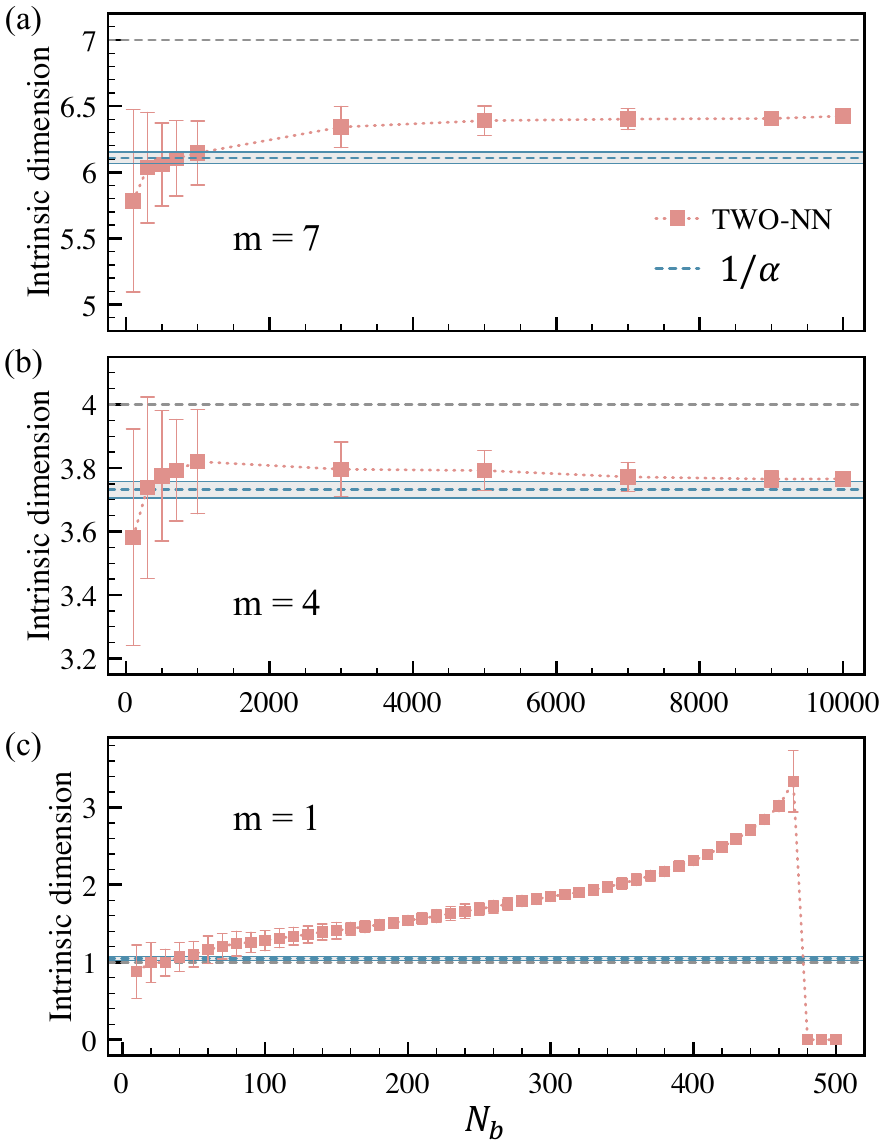}
    \caption{Dependencies of the intrinsic dimension on the number of unique samples $N_b$ for the synthetic datasets with (a) $m=7$, (b) $m=4$, and (c) $m=1$. The horizontal gray dashed lines correspond to the true predefined intrinsic dimension of the manifold ($m$). The blue dashed line denotes the $1/\alpha$ value calculated for the ranges presented in Section \textit{Estimating the Intrinsic Dimension}. The shaded areas around the blue dashed lines represent standard deviations of $1/\alpha$ calculated over 30 independent runs. For the TWO-NN method, the results are presented for a single run along with the corresponding standard deviations.
}
    \label{ID_N=500}
\end{figure}

To determine the intrinsic dimension of a dataset via the TWO-NN approach, one evaluates it at different sample sizes $N_b$ and tracks the behavior of the resulting dependence $d_{\rm {TWO-NN}}(N_b)$ \cite{TWO-NN}. The true intrinsic dimension is defined either as the saturation plateau value or as the value attained at the maximum number of unique samples when saturation is not reached. 
For instance, for our synthetic datasets with $N \approx 500$, the function $d_{\rm{TWO-NN}}(N_b)$ saturates for all dimensions $m \ge 2$, as illustrated in Fig.~\ref{ID_N=500}(a, b). The opposite behavior is observed for the one-dimensional case with $m=1$ (see Fig.~\ref{ID_N=500}(c)), where the estimated intrinsic dimension initially increases with $N_b$ and then sharply drops to zero 
near the maximum number of unique bitstrings.  In this case the condition $r_1=r_2$ holds for every sample and the slope of $-\ln[1 - F^{\rm emp}(\mu)](\ln(\mu))$ line is zero.

The intrinsic dimension $d_{\rm {TWO-NN}}$ can be successfully utilized to characterize the structural features of quantum states \cite{Dalmonte1} and serves to identify the most appropriate measurement basis for analyzing specific quantum systems \cite{Dalmonte3}. In some cases, evaluating the intrinsic dimension across different bases even facilitates the detection of 
quantum phase transitions. In this respect, the $d_{\rm TWO-NN}$ is closely related to the dissimilarity measure proposed recently by some of us~\cite{dissimilarity}. This quantity was designed to detect phase transitions in quantum systems and is based on the analysis of pattern scales presented in data.

\begin{figure}[!b]
    \includegraphics[width=\linewidth]{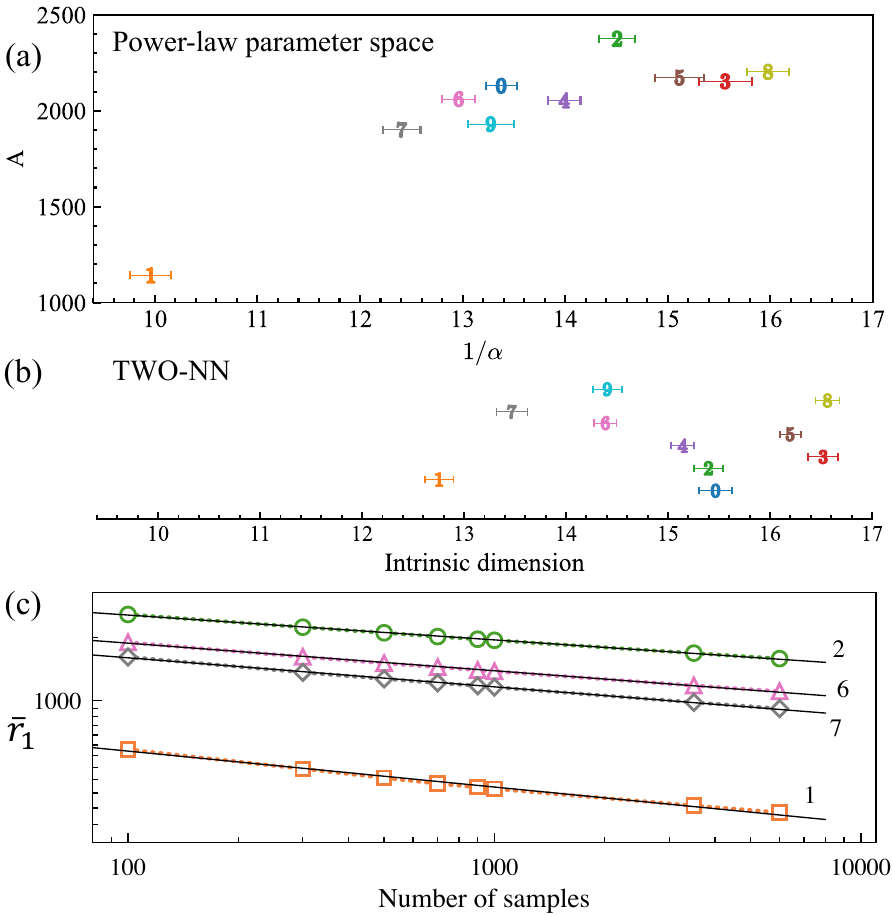}
    \caption{Low-dimensional representations of the digit classes from the MNIST dataset based on (a) the power-law parameters (fitted up to $N_b^{\rm max} = 6000$) and (b) the intrinsic dimension $d_{\rm {TWO-NN}}$ evaluated on $6000$ samples. The power-law parameters are additionally averaged over 30 independent runs to evaluate the standard deviations; error bars for $A$ are smaller than the symbol size. (c) Dependencies of $\bar{r}_1$ versus $N_b$ for selected digits from the MNIST dataset. The error bars are smaller than the symbol size, and the solid black lines show the corresponding power-law fits.
}
    \label{MNIST}
\end{figure}

\subsection{Appendix C: Tests on classical real-world data}
In this section, we discuss the potential application of our approach to real-world classical data, using the benchmark MNIST dataset obtained from the Keras Python library~\cite{Keras}. This dataset consists of $70,000$ $28 \times 28$ grayscale images of handwritten digits. To calculate $\bar r_1$, $r_1$ and $r_2$ between data points in the initial 784-dimensional space we employ the standard Euclidean distance with sample sizes chosen in the range $N_b\in[100, 6000]$. As can be seen from Fig.~\ref{MNIST}, $1/\alpha$ is generally lower than the intrinsic dimension $d_{\rm {TWO-NN}}$ evaluated at $N_b=6000$ (Fig.~\ref{MNIST}(b)). However, the hierarchy among the different digit classes remains remarkably consistent in both cases, which demonstrates the feasibility of applying our geometric framework to general datasets of non-quantum origin. 

Based on this representative example, one can conclude that using $A$, $\alpha$ and $d_{\rm {TWO-NN}}$ together may help to distinguish different datasets. According to the results presented in Fig.~\ref{MNIST}(b), the digits 3 and 8, 6 and 9 as well as 2 and 0 are indistinguishable based only on the $d_{\rm {TWO-NN}}$ value. At the same time, (see Fig.~\ref{MNIST}(a)), the classes for digits 6 and 9 lie in different regions with respect to the $A$ parameter, and the situation for 2's and 0's is even better. However, the profiles for 3's and 8's still slightly overlap, suggesting that alternative metrics may be required to distinguish these particular datasets. In contrast, the digits 5 and 3 overlap in terms of their $\alpha$ exponents and lie within the same range of $A$, but they remain clearly distinguishable based on their intrinsic dimension $d_{\rm {TWO-NN}}$.

Figure~\ref{MNIST}(c) shows that the power-law behavior of $\bar r_1(N_b)$ perfectly holds for the MNIST dataset, which supports claims concerning the broad applicability of the proposed approach. Moreover, it paves the way for using the $1/\alpha$ metric to evaluate the intrinsic dimension of datasets of non-quantum nature based solely on the expected distance to the first nearest neighbor. This is important, for instance, when duplication occurs or there are samples for which several closest neighbors are equally distant. As for the whole MNIST dataset, we obtained $1/\alpha\approx12.9$ and $d_{\rm TWO-NN}\approx14$ which is in line with previous estimations of the intrinsic dimension $d\in[7, 15.2]$ reported for the different versions of MNIST and various estimation schemes~\cite{ID_images1, ID_images2}. In our calculations, we used $N_b\in[100, 10000]$ and averaged the results over 50 independent random subsets.

\subsection{Appendix D: Details of $J_1-J_2$ model calculations}
In this section we present technical details concerning calculations of  the $J_1-J_2$ models and compare our results with previous works. For small-size systems (36 spins) we use exact diagonalization approach as realized in SpinED package \cite{spinED}. In turn, for larger systems one needs to employ an approximate scheme, such as variational neural quantum state (NQS) approach. We use the NetKet package realization \cite{Netket2019, Netket2022} of an attention-based vision transformer wave function \cite{ViT1,ViT2}, as it is one of the most promising architectures for solving large-scale frustrated magnetic models \cite{Carleo4242}. For all the system's sizes we use the same number of the training epochs (1500) and the same ViT architecture that comprises 4 layers, 12 attention heads and embedding dimension of 60, which results in $\sim 1.5 \times 10^5$ trainable parameters. To update the network parameters we use stochastic reconfiguration approach \cite{DeepViT} with the built-in \texttt{MetropolisExchange} sampling method. The corresponding number of samples is 4096. 

Tables \ref{J1J2-model} gives the ground state energies obtained for $J_1 - J_2$ Hamiltonian defined on supercells of different sizes. In the case of the $6 \times 6$ supercell we compare the ViT energies with the exact diagonalization results. The latter are in full agreement with previous works \cite{Poilblanc}. The largest deviations between ViT and ED results take place for critical values of $J_2$ from 0.5 to 0.6. At the same time the best agreement is observed for the N\'eel phase.  As an important benchmark for our non-frustrated results obtained at $J_2 = 0$ we consider the  Quantum Monte Carlo (QMC) method which provides accurate ground state energies for large-scale systems (up to  $96 \times 96$ spins in \cite{Sandvik_2026}). The relative errors of the ViT ground state energies with respect to the QMC ones are approximately equal to $3 \times 10^{-4}$ \% ($6 \times 6$ supercell), $8 \times 10^{-3}$ \% ($8 \times 8$) and  $1 \times 10^{-2}$ \% ($10 \times 10$). For the systems on the $8 \times 8$ and $10 \times 10$ supercells simulated by using $J_2 > 0$ we compare our ViT estimations of the ground state energy with best results (the lowest energies) available in the literature. From Table \ref{J1J2-model} one can see that for some values of $J_2$ we obtain a lower energy than known estimates \cite{QMBL}.

\begin{figure}[!t]
    \includegraphics[width=\linewidth]{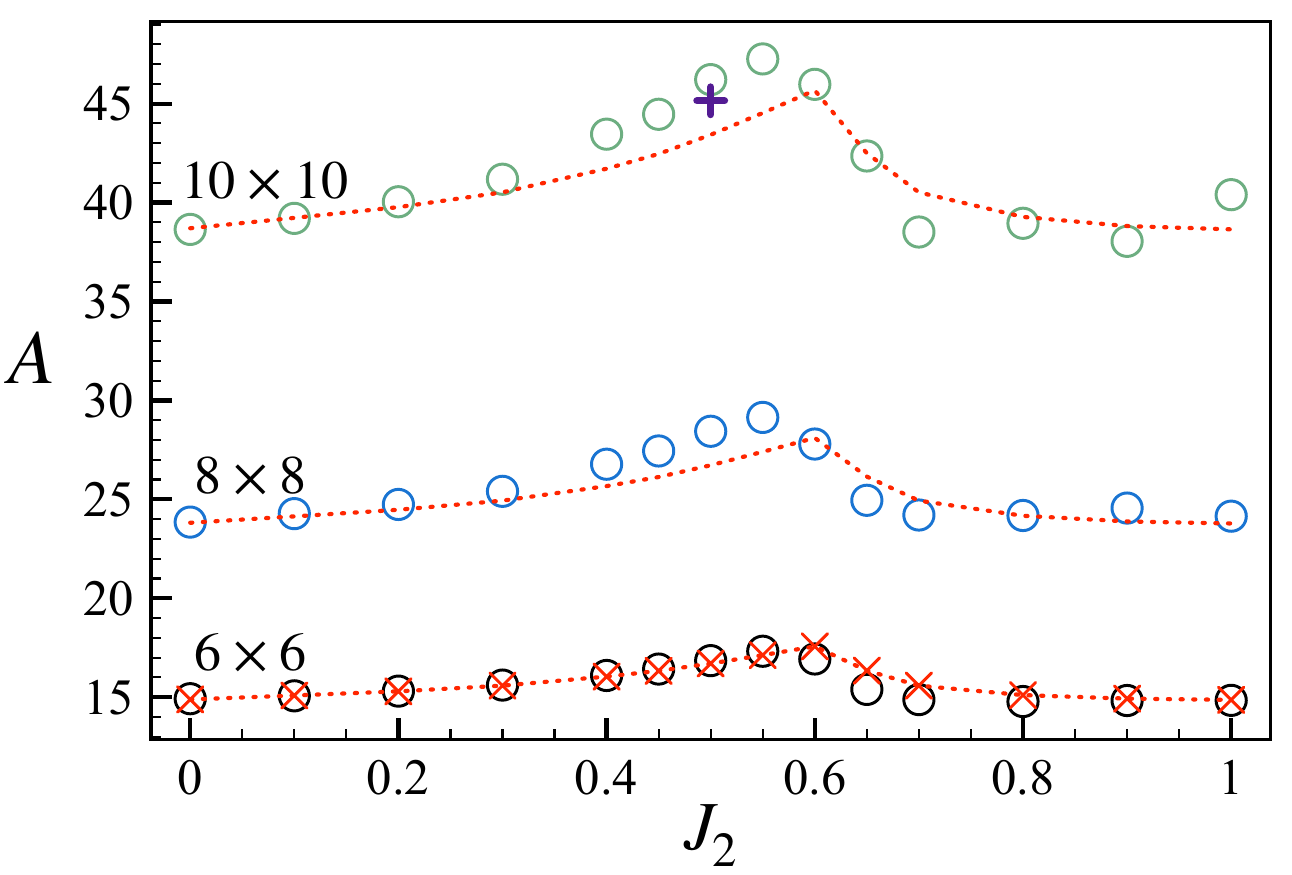}
    \caption{Power-law parameter, $A$ calculated for exact ED and approximated ViT ground states of the antiferromagnetic $J_1-J_2$ model. Black, blue and green circles correspond to the ViT results obtained after 1500 training epochs for $6 \times 6$, $8 \times 8$ and $10 \times 10$ supercells, respectively. Red crosses stand for exact diagonalization solutions of the $6 \times 6$ system. Red dotted lines denote ED data rescaled as described in the main text. The violet plus sign corresponds to the estimate based on pre-trained ViT data from \cite{DeepViT}. The error bars with respect to the averaging over non-overlapping datasets of $N_{b}^{\rm max}=2\cdot10^4$ unique outcomes are smaller than the symbol size.}
    \label{A_parameter}
\end{figure}

Fig.~\ref{A_parameter} complements Fig.\ref{Heis} presented in the main text and gives the $A$ parameters from the power-law fitting (Eq.~\eqref{power-law}) of the $\bar{r}_1 (N_b)$ functions calculated for the $J_1-J_2$ model of different sizes. For fitting with power law, Eq.~\eqref{power-law} we use the range $N_{b} \in [50, 1000]$. The ViT results denoted with color circles features maximum at 0.55, which agrees with the V-score metric.

\begin{table*}[hbt!]
    \centering
    \caption{Ground state energies of the $J_1 - J_2$ model obtained in this work with exact diagonalization (ED) and variational NQS approach utilizing vision transformer wave function (ViT). The comparison with the best results of previous works is also given. }
    \begin{tabular}{ | c | c | c | c | c | c | c | }
    \hline
            \multicolumn{7}{|c|}{ $J_1-J_2$ model} \\
        \hline
            \multirow{1}{4em}{\centering$J_2$} & \multirow{1}{8em}{\centering $6\times6$, ED } & \multirow{1}{8em}{\centering $6\times6$, ViT } &  \multirow{1}{8em}{\centering $8\times8$, ViT } & \multirow{1}{8em}{\centering $8\times8$} & \multirow{1}{8em}{\centering $10\times10$, ViT}  & \multirow{1}{8em}{\centering $10\times10$ }\\
            [1.5pt]
            \multirow{1}{4em}{\centering} & \multirow{1}{8em}{\centering (this work)} & (this work) & \multirow{1}{8em}{\centering (this work)} & \multirow{1}{8em}{\centering (previous works)} & \multirow{1}{8em}{\centering (this work)}  & \multirow{1}{8em}{\centering (previous works)}\\
            [1.5pt]
        \hline
        0.00 & -0.678872 & -0.678870 & -0.673434 & -0.67349005 [\onlinecite{Sandvik_2026}] & -0.671476 & -0.67155266[\onlinecite{Sandvik_2026}]\\ [1.5pt]
        0.10 & -0.638095 & -0.638078 & -0.633093 & & -0.631365 & - \\ [1.5pt] 
        0.20 & -0.599046 & -0.599044 & -0.594508 &  & -0.592982 & -0.592847 [\onlinecite{Chen}]\\ [1.5pt]
        0.30 & -0.562459 & -0.562453 & -0.558295 & & -0.556912 & -\\ [1.5pt]
        0.40 & -0.529744 & -0.529547 & -0.525513 & -0.525492 [\onlinecite{Imada}] & -0.524361 & -0.5240 [\onlinecite{Sorella}] \\ [1.5pt]
        0.45 & -0.515658 & -0.515281 & -0.510970 & -0.511117 [\onlinecite{Imada}] & -0.509661 & -0.51001 [\onlinecite{Sorella}]\\ [1.5pt]
        0.50 & -0.503809 & -0.502970 & -0.498115 & -0.498460 [\onlinecite{Imada}] & -0.496783 & -0.4976921 [\onlinecite{SSE_Heyl}]\\ [1.5pt]
        0.55 & -0.495177 & -0.493676 & -0.483143 & -0.48781 [\onlinecite{Imada}] & -0.485759 & -0.48693 [\onlinecite{Sorella}]\\ [1.5pt]
        0.60 & -0.493238 & -0.490755 & -0.481606 & & -0.477623 & -0.47839 [\onlinecite{Liang}]\\ [1.5pt]
        0.65 & -0.506590 & -0.506299 & -0.498312 & &-0.494672 & - \\ [1.5pt]
        0.70 & -0.529989 & -0.529884 & -0.522091 & &-0.519035 & -0.51889 [\onlinecite{Chen}]\\ [1.5pt]
        0.80 & -0.586485 & -0.585780 & -0.577310 & &-0.574130 & -0.57404 [\onlinecite{Chen}]\\ [1.5pt]
        0.90 & -0.649048 & -0.647371 & -0.637518 & &-0.634170 & - \\ [1.5pt]
        1.00 & -0.714356 & -0.712413 & -0.700508 & &-0.696641 & -0.69670 [\onlinecite{Chen}]\\ [1.5pt]
        \hline   
    \end{tabular}
    \label{J1J2-model}
\end{table*}

\end{document}